\documentclass[aps,nofootinbib,prd,eqsecnum,showpacs,showkeys,preprintnumbers,altaffilletter]{revtex4-1}

\usepackage[normalem]{ulem}
\usepackage{amsmath}
\usepackage{amssymb}

\usepackage{graphicx}
\usepackage{amsfonts}
\usepackage{color}
\usepackage{bm}
\usepackage{mathrsfs}
\usepackage{epstopdf}
\usepackage{url}
\usepackage{footnote}
\usepackage{appendix}

\usepackage{float}
\usepackage{enumerate}
\usepackage{lineno}

\usepackage{hyperref}

\usepackage{tabularx}

\makeatletter

\newcommand{\stkout}[1]{\ifmmode\text{\sout{\ensuremath{#1}}}\else\sout{#1}\fi}

\usepackage{booktabs}
\AtBeginDocument{
\heavyrulewidth=.08em
\lightrulewidth=.05em
\cmidrulewidth=.03em
\belowrulesep=.65ex
\belowbottomsep=0pt
\aboverulesep=.4ex
\abovetopsep=0pt
\cmidrulesep=\doublerulesep
\cmidrulekern=.5em
\defaultaddspace=.5em
}

\newcolumntype{L}[1]{>{\hsize=#1\hsize\raggedright\arraybackslash}X}%
\newcolumntype{R}[1]{>{\hsize=#1\hsize\raggedleft\arraybackslash}X}%
\newcolumntype{C}[1]{>{\hsize=#1\hsize\centering\arraybackslash}X}%

\newcommand*\patchAmsMathEnvironmentForLineno[1]{%
 \expandafter\let\csname old#1\expandafter\endcsname\csname #1\endcsname
 \expandafter\let\csname oldend#1\expandafter\endcsname\csname end#1\endcsname
 \renewenvironment{#1}%
   {\linenomath\csname old#1\endcsname}%
   {\csname oldend#1\endcsname\endlinenomath}}%
\newcommand*\patchBothAmsMathEnvironmentsForLineno[1]{%
 \patchAmsMathEnvironmentForLineno{#1}%
 \patchAmsMathEnvironmentForLineno{#1*}}%
\AtBeginDocument{%
\patchBothAmsMathEnvironmentsForLineno{equation}%
\patchBothAmsMathEnvironmentsForLineno{align}%
\patchBothAmsMathEnvironmentsForLineno{flalign}%
\patchBothAmsMathEnvironmentsForLineno{alignat}%
\patchBothAmsMathEnvironmentsForLineno{gather}%
\patchBothAmsMathEnvironmentsForLineno{multline}%
}

\definecolor{greenNew}{rgb}{0.15,0.75,0.0}

\newcommand{\BRparamA}{{w_{\textrm{br}}}}
\newcommand{\LRparamA}{{\Omega_\textrm{lr}}}
\newcommand{\LSBRparamA}{{\Omega_\textrm{lsbr}}}

\newcommand{\rad}{{\textrm{r}}}
\newcommand{\mat}{{\textrm{m}}}
\newcommand{\de}{{\textrm{d}}}
\newcommand{\ini}{{\textrm{ini}}}

\makeatother

\begin{document}

\title{Cosmological perturbations in an effective and genuinely phantom dark energy Universe}

\author{Imanol Albarran $^{1,2}$}
\email{imanol@ubi.pt}

\author{Mariam Bouhmadi-L\'{o}pez $^{3,4}$}
\email{{\mbox{mariam.bouhmadi@ehu.eus}}}

\author{Jo\~{a}o Morais $^{3}$ }
\email{jviegas001@ikasle.ehu.eus}

\date{\today}

\affiliation{
${}^1$Departamento de F\'{\i}sica, Universidade da Beira Interior, Rua Marqu\^{e}s D'\'Avila e Bolama 6200-001 Covilh\~a, Portugal\\
${}^2$Centro de Matem\'atica e Aplica\c{c}\~oes da Universidade da Beira Interior, Rua Marqu\^{e}s D'\'Avila e Bolama 6200-001 Covilh\~a, Portugal\\
${}^3$Department of Theoretical Physics University of the Basque Country UPV/EHU. P.O. Box 644, 48080 Bilbao, Spain\\
${}^4$IKERBASQUE, Basque Foundation for Science, 48011, Bilbao, Spain\\
}

\begin{abstract}

We carry out an analysis of the cosmological perturbations in general relativity for three different models which are good candidates to describe the current acceleration of the Universe. These three set-ups are described classically by perfect fluids with a phantom nature and represent deviations from the most widely accepted $\Lambda$CDM model. In addition, each of the models under study induce different future singularities or abrupt events known as (i) Big Rip, (ii) Little Rip and (iii) Little Sibling of the Big Rip. Only the first one is regarded as a true singularity since it occurs at a finite cosmic time. For this reason, we refer to the others as abrupt events.
With the aim to find possible footprints of this scenario in the Universe matter distribution, we not only obtain the evolution of the cosmological scalar perturbations but also calculate the matter power spectrum for each model.  We have carried the perturbations in the absence of any anisotropic stress and within a phenomenological approach for the speed of sound. We constrain observationally these models using several measurements of the growth rate function, more precisely $f\sigma_8$, and compare our results with the observational ones.

\end{abstract}



\maketitle

%
%

\section{Introduction}

Cosmology has made a long way on the last years with the impressive amount of observations and theoretical advancements. Yet, it still faces many challenging questions like the fundamental cause of the recent acceleration of the Universe, which was found with SNeIa observations almost twenty years ago \cite{Riess:1998cb,Perlmutter:1998np}, and afterwards confirmed by several types of cosmological and astrophysical observations (see for example \cite{Ade:2015rim} for a recent account on this issue).
The simplest approach which is in agreement with the current observations is to assume a cosmological constant that started recently  to dominate the late-time energy density budget of the Universe \cite{Ade:2015xua}. But then the issues \textit{of why is it so tiny?} and \textit{why this cosmological constant has begun to be important only right now?} have to be addressed as well (see for example: \cite{Sahni:1999gb,Carroll:2000fy,Weinberg:1988cp,Padmanabhan:2002ji}).
Another, equally important, issue is what happens if the cosmological constant is not quite constant? This has led to a great interest in exploring other possible scenarios to explain the late-time acceleration of the cosmos by invoking either an additional matter component in the Universe, which we name dark energy (DE) \cite{Tsujikawa:2010sc,Bamba:2012cp,AmendolaTsujikawa}, or by modifying appropriately the laws of gravity (for a recent account on this issue see, for example, \cite{CapozzielloFaraoni,Morais:2015ooa} and the extensive list of references provided therein).

We will focus on the third question: {\textit{what happens if the cosmological constant is not quite constant?}} More precisely, we will address this question on the framework of the cosmological perturbations and for some DE models whose equations of state; i.e., the ratio between its pressure and its energy density, deviate slightly from the one corresponding to a cosmological constant. Before proceeding let us remind the following well known fact: if the equation of state (EoS) of DE deviates from -1, the Universe fate might be quite different from the one corresponding to an empty de Sitter Universe. In particular, if the equation of state of dark energy is smaller than -1, i.e., DE is apparently (at least from an effective point of view) not fulfilling the null energy condition, several future singularities or abrupt events might correspond to the cosmic doomsday of the Universe. Amazingly, some of these models are in accordance with current data \cite{Jimenez:2016sgs}.

On the other hand, theory of cosmological perturbations is a cornerstone of nowadays cosmology. It provides us with a theoretical framework which allows us to determine, for example, the CMB predicted from an early inflationary era or compute the matter power spectrum and the growth rate of matter in order to make a comparison with the observational results. In addition, it allows us to compute the evolution and possible clustering of DE perturbations and investigate their effect on the growth of dark matter (DM).
Even though no perturbations of DE have so far been detected, and are in fact absent in the $\Lambda$CDM model, the existence of a great number of experiments aiming to probe the physics of the late-Universe, like the Dark Energy Survey \cite{Abbott:2016ktf} and the Euclid mission \cite{Amendola:2016saw}, suggests that a thorough study and characterization of such effects can be proven to be important to understand the nature of this mysterious fluid that drives the acceleration of the Universe. {With this mindset, on this work we analyse the perturbative effects of phantom DE models%
\footnote{There are some promising phantom dark energy models \cite{Deffayet:2010qz,Pujolas:2011he} which are free from ghosts and gradient instabilities (see also Ref.~\cite{Easson:2016klq}).}%
 and look for observational fingerprints that could be used as a mean to favour or disregard such models.}

There are three genuinely phantom DE fates; i.e., which happens if and only if a phantom DE component is present:
\begin{itemize}
\item Big Rip (BR) singularity: It happens at a finite cosmic time with an infinite scale factor where the Hubble parameter and its cosmic time derivative diverge \cite{Starobinsky:1999yw,Caldwell:2003vq, Caldwell:1999ew,Carroll:2003st,Chimento:2003qy,Dabrowski:2003jm, GonzalezDiaz:2003rf,GonzalezDiaz:2004vq}.
\item Little Rip (LR): This case corresponds to an abrupt event, i.e., it is not strictly speaking a future space-time singularity, as it takes place at an infinite cosmic time \cite{Ruzmaikina}. The radius of the Universe, the Hubble parameter and its cosmic time derivative all diverge at an infinite cosmic time \cite{Nojiri:2005sx,Nojiri:2005sr,
Ruzmaikina,Stefancic:2004kb,BouhmadiLopez:2005gk,Frampton:2011sp,
Bouhmadi-Lopez:2013nma,Brevik:2011mm}.
\item Little Sibling of the Big Rip (LSBR): This behaviour again corresponds to an abrupt event rather than a future space-time singularity. At this event, the Hubble rate and the scale factor blow up but the cosmic derivative of the Hubble rate does not \cite{Bouhmadi-Lopez:2014cca,Morais:2016bev}. Consequetly, this abrupt event takes place at an infinite cosmic time where the scalar curvature diverges.
\end{itemize}

These three cases share in common the fact that in the (far) future all the structures in the Universe would be ripped apart in a finite cosmic time \cite{Frampton:2011sp,Bouhmadi-Lopez:2014cca}. The classical singular asymptotic behaviour of these dark energy models has led to a quantum cosmological analysis of these setups \cite{Dabrowski:2006dd,Albarran:2015tga,Albarran:2015cda, Albarran:2016ewi,Bouhmadi-Lopez:2016dcf}. In these works, it was concluded that once the Universe enters in a genuinely quantum phase; i.e., where coherence and entanglement effects are important, the Universe would evade a doomsday \textit {\`{a} la rip}. This applies even to the smoother version of these singular behaviours corresponding to a LSBR \cite{Albarran:2016ewi} (see also \cite{Kamenshchik:2007zj,BouhmadiLopez:2009pu,Kamenshchik:2012ij,Kamenshchik:2013naa,Bouhmadi-Lopez:2013tua}).

In this paper, we will analyse the cosmological perturbations of DE models that induce a BR, LR or LSBR. While the background analysis of the phantom DE scenario has been widely analysed, this has not been the case of its cosmological perturbations \cite{Kunz:2006wc,Balcerzak:2012ae,Astashenok:2012iy,Denkiewicz:2014aca,Denkiewicz:2015nai}. 
In \cite{Balcerzak:2012ae,Denkiewicz:2014aca,Denkiewicz:2015nai} a kinematical approach was assumed, i.e., a dependence of the scale factor as a function of the cosmic time was considered for FLRW Universes with future singularities at a finite cosmic time. 
Within this setup and using approximated equations for the growth of the perturbations at late-time, the authors obtained the DM and DE perturbations \cite{Denkiewicz:2014aca}. Furthermore, in \cite{Balcerzak:2012ae,Denkiewicz:2015nai}, DE perturbations are disregarded and only the growth rate of matter perturbations is calculated. In this work, we will rather assume a dynamical model, i.e., we assume a given EoS for DE. This is the approach employed in Ref \cite{Astashenok:2012iy}, where the future behaviour of the linear scalar perturbations is presented for a type of model that, depending on the value of the parameters, can lead to a BR or a Big freeze singularity \cite{BouhmadiLopez:2006fu}. In our analysis, we use the full theory of linear perturbation to study how the perturbations of DM and DE, as well as, the gravitational potential evolve for a range of different scales. Our numerical integrations start from well inside the radiation era and continue till the far future. In fact, in order to see the behaviour of the phantom DE models, we extend our numerical calculations till the Universe is roughly $e^{12}$ times larger than at present, i.e., roughly $z\thicksim-1$.
In the perturbative analysis carried out we (i) disregard any anisotropic stress tensor, (ii) consider the DE perturbations to be non-adiabatic and (iii) describe this non-adiabaticity within a phenomenological approach for the speed of sound. On the other hand, we disregard the contribution of neutrinos as a first approach where we do not use a more advanced Boltzman code such as CAMB \cite{CAMB} or CLASS \cite{Lesgourgues:2011re}.

The paper is organised as follows:
In Sect.~\ref{review}, we briefly review DE models that induce a BR, LR or LSBR; i.e., we review perfect the fluids that can describe DE from a phenomenological point of view and leading to the above mentioned singularity or abrupt events.
In Sect.~\ref{Perturbed equations}, we present the framework for studying the scalar linear perturbations. The evolution equations obtained employ no approximations and are therefore valid for all relevant modes and redshifts, as long as the linear theory is itself valid.
We present our numerical results in Sect.~\ref{results}, where we show the evolution of different perturbed quantities related to DM and DE. We present as well the matter power spectrum for the different models. We equally constrain these models using several measurements of the growth rate function, more precisely $f\sigma_8$. 
Finally, in Sect.~\ref{conclusions}, we present our conclusions.

%
%

\section{Background models}
\label{review}

In this section, we briefly review the different models that, at the background level, lead to distinct future cosmological abrupt events: (i) Big Rip (BR), (ii) Little Rip (LR) and (iii) Little Sibling of the Big rip (LSBR). For each of these models, we begin by presenting an EoS for DE that can originate such genuinely phantom abrupt events in the future, while ensuring that the background evolution follows closely that of $\Lambda$CDM until the present time. These models should be interpreted as an effective description of a more fundamental field, therefore, even though at the background level they might be defined by a barotropic fluid, the same should not be assumed at the perturbative level. In fact, as we will discuss below, in order to avoid non-physical instabilities we will explicitly break the adiabaticity of the DE perturbations. Bearing in mind, from now on, the approach that we will follow, we next describe the effective background models that we will contemplate.

Let us consider a homogeneous and isotropic Universe described by the Friedmann-Lema{\^i}tre-Robertson-Walker (FLRW) metric:
\begin{align}
	ds^2=-dt^2+a^2\left(t\right)\left[\frac{dr^2}{1-kr^2}+r^2d\theta^2+r^2\sin^{2}\theta d\varphi^2\right]
	\,,
\end{align}
where $a(t)$ is the scale factor and $k=-1,0,1$ for open, flat and closed spatial geometry, respectively. We will focus on the spatially flat case $(k=0)$, for which the Friedmann and Raychaudhuri equations read
\begin{align}
\label{Friedman}
	H^2=\frac{8\pi G}{3}\rho
	\,,
\end{align}
\begin{align}
\label{conservback}
	\dot{H}=-4\pi G\left(\rho +p\right)
	\,.
\end{align}
Here, $H$ is the Hubble parameter, a dot represents a derivative with respect to the cosmic time, $t$, $G$ is the cosmological constant and $\rho$ and $p$ are the total energy density and pressure of all the matter content of the Universe.
In this work, we will consider the Universe to be filled by radiation, dust (cold dark matter and baryons), and DE. As such, we can decompose $\rho$ and $p$ as
\begin{align}
\label{back1}
	\rho=\rho_{\rad}+\rho_{\mat}+\rho_\de 
	\qquad \textrm{and} \qquad 
	p=p_{\rad}+p_{\mat}+p_\de
	\,,
\end{align}
where $\rho_{\rad}$, $\rho_{\mat}$, and $\rho_\de$ correspond to the energy density of radiation, matter (cold dark matter and baryons) and DE. Similarly, $p_{\rad}$, $p_{\mat}$, and $p_\de$ are the pressure of radiation $(p_{\rad}=1/3\rho_{\rad})$, matter $(p_{\mat}=0)$, and DE $(p_\de=w_\de\rho_\de)$.
We will not take into account interactions between the individual matter components. Consequently, each fluid $A=\rad$,$\mat$,$\de$ verifies the usual conservation equation:
\begin{align}
\label{conservindiv}
	\dot{\rho}_A + 3H\left(\rho_A+p_A\right)=0
	\,.
\end{align}
For latter convenience, we define the fractional energy density of the individual mater components as
\begin{align}
\label{back2}
	\Omega_{\rad}=\frac{\rho_{\rad}}{\rho}
	\,, 
	\qquad 
	\Omega_{\mat}=\frac{\rho_{\mat}}{\rho}
	\,,
	\qquad
	\Omega_\de=\frac{\rho_\de}{\rho}
	\,,
\end{align}
and the individual parameters of EoS
\begin{align}
	w_{\rad}=\frac{p_{\rad}}{\rho_{\rad}}=\frac{1}{3} 
	\,, 
	\qquad 
	w_{\mat}=\frac{p_{\mat}}{\rho_{\mat}}=0 
	\,,
	 \qquad w_\de=\frac{p_\de}{\rho_\de}
	 \,.
\end{align}
The DE parameter of EoS, $w_\de$, will be fixed later for each individual model. From (\ref{back1}) and (\ref{back2}) we can obtain the total parameter of EoS, $w$, from the individual $w_A$ as:
\begin{align}
\label{backwtotal}
	w\equiv\frac{p}{\rho}=\Omega_{\rad} w_{\rad}+\Omega_{\mat} w_{\mat}+\Omega_\de w_\de
	\,.
\end{align}

In the following subsections, we present the three models which will be studied in this work. From this point onward a $0$-subscript denotes the present value of a given quantity. On the other hand, we will denote with a $*$-subscript, a point in the future evolution of the Universe, $a\left(t_{*}\right)=a_{*}$, where DE totally dominates over the matter, i.e. $\rho_\mat\left(a_{*}\right)\ll\rho_\de\left(a_{*}\right)$.

%
%

\subsection{BR singularity: model (i)}
\label{BR}

A BR singularity \cite{Starobinsky:1999yw,Caldwell:1999ew,Caldwell:2003vq,Carroll:2003st, Chimento:2003qy,Dabrowski:2003jm,GonzalezDiaz:2003rf,GonzalezDiaz:2004vq} can be induced by a perfect fluid whose  EoS  parameter,  $w_\de$, is constant and smaller than $-1$: 
\begin{align}
	\label{eosbr}
	p_\de = \BRparamA \rho_\de
	\,.
\end{align}
Therefore, using the conservation equation (\ref{conservindiv}), the energy density evolves with the scale factor as
\begin{align}\label{ehobr}
	\rho_\de\left(a\right) =\rho_{\de0} \,\left(\frac{a}{a_0}\right)^{-3\left(1+\BRparamA\right)}
	\,.
\end{align}
Finally, the asymptotic evolution of the scale factor is 

\begin{align}\label{clatraybr}
	a\left(t\right)\sim
	a_0\left[\frac{3}{2}\left|1+\BRparamA\right|H_0\sqrt{\Omega_{\de0}}\left(t_\textrm{s}-t\right)\right]^{\frac{2}{3\left(1+\BRparamA\right)}
	\,.
	}
\end{align}
 where $t_s$ corresponds to the time of the singularity.
In this kind of future singularity, the scale factor, the Hubble parameter and its cosmic time derivatives blow up at a finite cosmic time. On the rest of the work, for the BR model we fix the free parameters of the model to the best fit in accordance with the Planck data for wCDM model \cite{wikiesa}: $w_\de=-1.019$, $\Omega_{\textrm{m0}}=0.306$ and $H_0=68.1 \ \textrm{km} \ \textrm{Mpc}^{-1} \textrm{s}^{-1}$ (please, cf. page 687 of \cite{wikiesa}). 

%
%

\subsection{LR abrupt event: model (ii)}
\label{LR}

The case of a LR \cite{Nojiri:2005sx,Nojiri:2005sr, Ruzmaikina,Stefancic:2004kb,BouhmadiLopez:2005gk,Frampton:2011sp, Bouhmadi-Lopez:2013nma,Brevik:2011mm} can be caused by a perfect fluid whose EoS fulfils \cite{Nojiri:2005sx,Stefancic:2004kb}
\begin{align}
\label{eoslr}
	p_\de=-\left(\rho_\de +\mathcal{B} \rho_\de^{\frac{1}{2}}\right)
	\,,
\end{align}
where the constant $\mathcal{B}$ is positive and has dimensions of inverse length square. 
Applying the conservation equation (\ref{conservindiv}), the energy density in terms of the scale factor reads
\begin{align}
\label{eholr}
	\rho_\de\left(a\right) =\rho_{\de0} \left[
		1
		+
		\frac{3}{2} \sqrt{\frac{\LRparamA}{\Omega_\textrm{d0}}} \ln\left(\frac{a}{a_0}\right)
	\right]^2
	\,,
\end{align}
where we have introduced the dimensionless parameter $\LRparamA \equiv (8\pi G)/(3H_0^2) \mathcal{B}^2$.
Finally, the asymptotic future evolution of the scale factor can be written as \cite{Frampton:2011sp}
\begin{align}
\label{clatraylr}
	a\left(t\right) \sim 
	a_0 \exp\left[\exp\left(\frac{3}{2}\sqrt{\LRparamA}H_0 t\right)\right]
	\,,
\end{align}
 In this kind of abrupt event, the scale factor, the Hubble parameter and its cosmic time derivatives blow up at an infinite cosmic time.  
For our later numerical calculations and as a guideline we fix $\Omega_{\textrm{m0}}=0.306$ and $H_0=68.1 \ \textrm{km} \ \textrm{Mpc}^{-1} \textrm{s}^{-1}$ as given by Planck \cite{Ade:2015xua,wikiesa} and $\LRparamA=1.2898\times 10^{-4}$.{ The value of the parameter $\LRparamA$ is chosen such that at the start of our numerical calculations the corresponding EoS parameters is equal to the one given by the model (i), i.e. $w_{\textrm{lr}}\left(a_{\textrm{ini}}\right)=\BRparamA$.
\footnote{\label{foot2} 
This equality implies that initially the DE perturbations $\delta_\de$ of the two models are also equal if the condition \eqref{initcond1} is imposed.
}

%
%

\subsection{LSBR abrupt event: model (iii)}
\label{LSBR}

The LSBR can be induced by a perfect fluid whose EoS deviates from that of a cosmological constant as \cite{Bouhmadi-Lopez:2014cca}
\begin{align}
\label{eoslsbr}
	p_\de=-\left(\rho_\de +\frac{\mathcal{A}}{3}\right)
	\,,
\end{align}
where $\mathcal{A}$ is a positive constant and whose dimensions are length to the fourth power. Making use of the conservation equation (\ref{conservindiv}), the energy density on this case reads
\begin{align}
\label{eholsbr}
	\rho_\de\left(a\right) =\rho_{\de0}
	\left[
		1
		+\frac{\LSBRparamA}{\Omega_{d0}}\ln\left(\frac{a}{a_0}\right)
	\right]
	\,,
\end{align}
where  we have introduced the dimensionless parameter $\LSBRparamA \equiv (8\pi G)/(3H_0^2) \mathcal{A}$. Finally, the future asymptotic growth of the scale factor with respect to the cosmic time can be written as \cite{Bouhmadi-Lopez:2014cca}
\begin{align}
\label{clatraylsbr}
	a\left(t\right)
	\sim
	 a_0 \exp\left(\frac{1}{4}\LSBRparamA H_0^2 t^2\right)
	\,.
\end{align}
In this type of abrupt event, the scale factor and the Hubble parameter blow up at infinite cosmic time. However, the cosmic time derivative of the Hubble parameter remains constant.
For our later numerical calculations and as a guideline we fix $\Omega_{\textrm{m0}}=0.306$ and $H_0=68.1 \ \textrm{km} \ \textrm{Mpc}^{-1} \textrm{s}^{-1}$ as given by Planck \cite{Ade:2015xua,wikiesa} and $\LSBRparamA=2.2134\times 10^{-2}$.{ The value of the parameter $\LSBRparamA$ is chosen such that at the start of our numerical calculations the corresponding EoS parameters is equal to the one given by the model (i), i.e. $w_{\textrm{lsbr}}\left(a_{\textrm{ini}}\right)=\BRparamA$.
\footnote{See footnote \ref{foot2}.}
}

\subsection{Comparing these models}

Aside from the definition of the BR, LR and LSBR given in the introduction, a few words are in order to compare the models we analyse in this work from a background point of view.
All the models presented above can be seen as a deviation from $\Lambda$CDM which can be recovered by setting $w_d=-1$ on the first model, $\mathcal{B}=0$ on the second model and $\mathcal{A}=0$ on the third model. Despite this apparent similarity with $\Lambda$CDM, they are characterised by a DE EoS satisfying $w < -1$, so they correspond to phantom models whose end state is drastically different from the de Sitter behaviour of a cosmological constant dominated Universe.
In all these cases, the Universe is not only accelerating but super accelerating asymptotically. This fact leads the universe unavoidably to unzip itself; i.e., all the bounded structures within it will be destroyed.
As can be seen from the asymptotic expansion of the scale factor $a(t)$ in Eq.~\eqref{clatraylr}, the BR is a true singularity as it would take place at a finite cosmic time from now. In addition, the geodesics cannot be extended beyond that point \cite{FernandezJambrina:2004yy}. On the other hand, the LR is more virulent than the LSBR as can be seen from Eqs.~\eqref{clatraylr} and \eqref{clatraylsbr}, although both of them would happen at an infinite time from now. 

In a FLRW background a phantom perfect fluid can in principle be described through a phantom scalar field, i.e., a minimally coupled scalar field with the opposite sign for its kinetic term \cite{Caldwell:1999ew}. In particular, this statement applies to the models we are considering. While a detailed study of this equivalence is not the purpose of this work, in the Appendix~\ref{Mapping to a phantom scalar field} we briefly explore the phantom scalar field model that could describe the phantom models (i), (ii) and (iii).

We will next analyse the behaviour of these models within the standard framework for the cosmological perturbations. As a first approach and in the rest of the work, we will disregard any anisotropic stress tensor and consider that the DE perturbations are non-adiabatic. As we will show, the second supposition is crucial to get a right description of the matter power spectrum. In addition, and as a matter of simplicity, the non-abiabaticity will be described within a phenomenological approach rather than in a more fundamental scope able to describe unequivocally and realistically the speed of sound. This issue is discussed in Section~\ref{speedsound_de}.

%
%

\section{Perturbed equations}
\label{Perturbed equations}

In this section, we review theory of linear perturbations for multi-fluid components. We choose the Newtonian gauge and work with the corresponding gauge invariant perturbation quantities. For a FLRW Universe, the perturbed line element is \cite{KurkiSuonio,DanielBaumann}
\begin{align}
\label{lineelem}
	ds^{2}=a^2\left[-\left(1+2\Phi\right)d\eta^2+\left(1-2\Psi\right)\delta_{ij}dx^idx^j\right]
	\,,
\end{align}
where $\eta$ is the comoving time, $d\eta=(1/a)dt$, a latin index denote purely spatial coordinates, and $\Psi(\eta,x^{i})$ and $\Phi(\eta,x^{i})$ are the gauge invariant Bardeen potentials \cite{Bardeen:1980gi}. The transformation rule $\dot{\{\,\}}=(1/a)\{\,\}'$, where a prime represents a derivative with respect to the conformal time, allows us to write $H$ and $\dot{H}$ in terms of the conformal Hubble parameter, $\mathcal{H}\equiv a'/a$, and its derivative, $\mathcal{H}'$, as
\begin{align}
\label{contime2}
	H=\frac{1}{a}\mathcal{H}
	\,, 
	\qquad \dot{H}=\frac{1}{a^2}\left(\mathcal{H}'-\mathcal{H}^2\right)
	\,.
\end{align}

Drawing from the line element (\ref{lineelem}), the inverse of the metric tensor can be obtained applying a Taylor expansion up to first order. Once we define the Christoffel symbols, we can compute the perturbation of the Ricci tensor, $\delta R_{\mu\nu}$, and of the curvature scalar, $\delta R$, in order to obtain the perturbed Einstein tensor $\delta G_{\mu\nu}\equiv \delta R_{\mu\nu} - \frac{1}{2}\delta^\mu_\nu \delta R$. In addition, the perturbed Einstein equations read
\begin{align}
\label{einstten}
	\delta G^{\mu}_{\phantom{a}\nu} 
	=8\pi G \delta T^{\mu}_{\phantom{a}\nu}
	\,,
\end{align}
where $\delta T^{\mu}_{\phantom{a}\nu}$ is the linear perturbation of the total energy momentum tensor. The individual components of Eq. (\ref{einstten}) can be written as \cite{KurkiSuonio,DanielBaumann}
\begin{align}\label{deltaT1}
\begin{split}
	3\mathcal{H}\left(\Psi' + \Phi\mathcal{H}\right) - \nabla^2 \Psi
	=&~4\pi G a^2\delta T_{\phantom{a}0}^{0}
	\,,\\
	- \left(\Psi' + \mathcal{H}\Phi\right)_{,i}
	=&~4\pi G a^2\delta T_{\phantom{a}i}^{0}
	\,, \\
	\Psi^{\prime\prime} 
	+ 2\mathcal{H}\left(\Phi'+ 2\Psi'\right)
	+ 2\Phi\left(2\mathcal{H}'+\mathcal{H}^2\right)\Phi 
	+ \frac{2}{3}\nabla^2\left(\Phi-\Psi \right)
	=&~\frac{4\pi G}{3} a^2\delta T_{\phantom{a}i}^{i}
	\,, \\
	\left(\Phi-\Psi\right)_{,ij}
	=&~8\pi G a^2\delta T_{\phantom{a}j}^{i}
	\,, 
	\qquad \qquad (i\neq j)\,.
\end{split}
\end{align}

The $\delta T^{\mu}_{\phantom{a}\nu}$ on the right hand side (rhs) of (\ref{einstten}) is the sum of the perturbations of the energy momentum tensor of radiation, $\delta T^{\mu}_{\rad\, \nu}$, non-relativistic matter (cold dark matter and baryons), $\delta T^{\mu}_{\mat \, \nu}$, and DE, $\delta T^{\mu}_{\de \, \nu}$. For each fluid, we can write the individual components of $\delta T^{\mu}_{A \, \nu}$ ($A=\rad,\mat,\de$) as 
\begin{align}\label{deltaT2}
\begin{split}
\delta T_{A \, 0}^{0}=&-\delta \rho_{A}
\,,\\
\delta T_{A \, 0}^{i}=&-\left(p+\rho\right)\partial^{i}v_{A}
\,, \\
\delta T_{A \, i}^{0}=&\left(p+\rho\right)\partial_{i}v_{A}
\,, \\
\delta T_{A \, j}^{i}=& \ \delta p_{A} \ \delta_{j}^{i}+\Pi_{A\,j}^{i}
\,,
\end{split}
\end{align}
where $\delta\rho_A$, $\delta p_A$, $v_A$, and $\Pi_{A\,j}^{i}$ are, respectively, the perturbation of the energy density, the perturbation of the pressure, the peculiar velocity potential and the anisotropic stress tensor of the fluid $A$. As a first approximation we consider that none of the fluids introduce anisotropies at the linear level of scalar perturbations. Therefore, from this point onward we will set $\Pi_{A\,j}^{i}=0$. From Eq.~\eqref{deltaT1}, we find that this implies the equality of the metric potentials $\Psi=\Phi$. Replacing this equivalence and Eqs.~\eqref{deltaT2} in the first three equations of~\eqref{deltaT1}, we obtain \cite{KurkiSuonio,DanielBaumann}
\begin{align}\label{GTequations2}
\begin{split}
	3\mathcal{H}\left(\mathcal{H}\Psi+\Psi'\right) -\nabla^{2}\Psi&= -4\pi Ga^{2}\delta\rho
	\,, \\
	\nabla^{2}\left(\mathcal{H}\Psi+\Psi'\right)&=-4\pi G a^{2}\left(\rho+p\right)\nabla^2 v
	\,,\\
	\Psi^{\prime\prime}+3\mathcal{H}\Psi'
+\Psi\left(2\mathcal{H}'+\mathcal{H}^{2}\right)&=4\pi G a^2\delta p
	\,,
\end{split}
\end{align}
where we have introduced the total energy density perturbation, $\delta \rho$, total pressure perturbation, $\delta p$, and total velocity potential, $v$. These can be written in terms of the individual fluid variables in Eqs.~\eqref{deltaT2} through the relations
\begin{align}\label{partialtotalrel}
	\delta \rho =~ \sum_{A}\delta\rho_A
	\,,
	\qquad
	\delta p =~ \sum_A\delta p_A
	\,,
	\qquad
	v =&~ \sum_A\frac{1+w_A}{1+w}{\Omega_A}v_A
	\,.
\end{align}
Following \cite{Bean:2003fb,Valiviita:2008iv}, we decompose the pressure perturbation as%
\footnote{%
The re-derivation of this expression is presented in the Appendix~\ref{AppA}.%
}
\begin{align}
	\label{pnew}
	\delta p_A=c_{sA}^2\delta\rho_A-3\mathcal{H}\left(1+w_A\right)\left(c_{sA}^2-c_{aA}^2\right)\rho_A v_A
	\,,
\end{align} 
where
\begin{align}\label{newvar0}
	c_{sA}^{2}=\left.\frac{\delta p_A}{\delta\rho_{A}}\right\vert_\textrm{r.f.}
	\,,
	\qquad 
	c_{aA}^2=\frac{p_A'}{\rho_A'}
	\,,
\end{align}
are, respectively, the effective squared speed of sound, defined in the rest frame (r.f.), of the fluid and the adiabatic speed of sound. In the following analysis, we will replace the energy density perturbation $\delta\rho_A$ by the fractional energy density perturbation $\delta_A=\delta\rho_A/\rho_A$. The total perturbation $\delta$ can be obtained from Eq.~\eqref{partialtotalrel} and reads
\begin{align}\label{deltaOmega}
	\delta= \sum_A \frac{\rho_A}{\rho}\delta_A = \sum_A \Omega_A\delta_A
	\,.
\end{align}

The perturbed conservation equations of the energy-momentum tensor for each fluid read:
\begin{align}\label{conserveq}
	\nabla_{\mu}\delta T_{A\,\nu}^{\mu}+\delta\Gamma_{\mu\alpha}^{\mu}T_{A\,\nu}^{\alpha}-\delta\Gamma_{\mu\nu}^{\alpha}T_{A\,\alpha}^{\mu}=0
	\,,
\end{align} 
where $\delta\Gamma_{\mu\nu}^{\alpha}$ is the perturbation of the Christoffel symbol and $T_{A\nu}^{\alpha}$ is the background value of the energy momentum tensor. Using Eqs. \eqref{deltaT2}, \eqref{pnew}, \eqref{newvar0}, and \eqref{deltaOmega}, we can compute the temporal and spatial components of Eq.~\eqref{conserveq} and obtain the evolution equations for the fractional energy density perturbation $\delta_A$ and the velocity potential $v_A$
\begin{align}\label{seteq1ncor}
\begin{split}
	&\delta_{A}'=
	3\mathcal{H}\left(w_A-c_{sA}^2\right)\delta_A
	+\left(1+w_A\right)\left[9\mathcal{H}^2\left(c_{sA}^2-c_{aA}^2\right)-\nabla^2\right]v_A
	+ 3\left(1+w_A\right)\Psi'
	\,,\\
&v_{A}'=\left(3c^2_{sA}-1\right)\mathcal{H}v_A-\frac{c^{2}_{sA}}{1+w_A}\delta_A - \Psi
	\,.
\end{split}
\end{align}
In summary, from the perturbed Einstein equation, we obtain Eqs.~\eqref{GTequations2} which relate the metric perturbations to the total perturbed matter quantities. On the other hand, from the perturbed conservation equations we obtain Eqs.~\eqref{seteq1ncor} which dictate the dynamics for the individual energy density and velocity perturbations.

In order to study the evolution of the linear perturbations, we conveniently apply a Fourier transformation, where we decompose a given function $\psi(\eta,\bold{x})$ into its Fourier components $\psi_k(\eta)$ as
\begin{align}\label{psifourier}
	\psi(\eta,\bold{x})=\frac{1}{\left(2\pi\right)^{3/2}}\int e^{-i{\bold{k} \cdot \bold{x}}}\psi_k(\eta) \ d^3\bold{k}
	\,.
\end{align}
 Therefore, for practical purposes, we make the substitution $ \nabla^{2}\rightarrow -k^{2}$ in all the evolution equations. On the other hand for our numerical calculations, we will apply the following change of variable:
\begin{align}
\label{varchange}
	x\equiv \ln\left(a\right)
	\,,
	\qquad
	\left\{\right\}'=\left\{\right\}_x \mathcal{H}
	\,,
	\qquad
	\left\{\right\}^{\prime\prime}=\left\{\right\}_{xx} \mathcal{H}^2+\left\{\right\}_x \mathcal{H}'
	\,,
\end{align}
where the subscript $x$ denotes a derivative with respect to $x$. By applying the Fourier decomposition \eqref{psifourier} and the transformation \eqref{varchange} to the sets of equations \eqref{GTequations2} and \eqref{seteq1ncor}, we obtain the evolution equations for each mode of the energy density and velocity perturbations of radiation, dust and dark energy
\begin{align}\label{seteq1each}
\begin{split}
	\left(\delta_{\rad}\right)_x &=\frac{4}{3}\left(\frac{k^2}{\mathcal{H}}v_{\rad}+3\Psi_x\right)
	\,,\\
	\left(v_{\rad}\right)_x &=-\frac{1}{\mathcal{H}}\left(\frac{1}{4}\delta_{\rad}+\Psi\right)
	\,,\\
	\left(\delta_{\mat}\right)_x &=\left(\frac{k^2}{\mathcal{H}}v_{\rad}+3\Psi_x\right)
	\,,\\
	\left(v_{\mat}\right)_x &=-\left(v_{\mat}+\frac{\Psi}{\mathcal{H}}\right)
	\,,\\
	\left(\delta_{\de}\right)_x &=\left(1+w_\de\right)\left\{\left[\frac{k^2}{\mathcal{H}} + 9\mathcal{H}\left(c_{s\de}^2-c_{a\de}^2\right) \right]v_\de+3\Psi_x\right\}+3\left(w_\de-c_{s\de}^2\right)\delta_\de
	\,,\\
	\left(v_{\de}\right)_x &=-\frac{1}{\mathcal{H}} \left(\frac{c_{s\de}^2}{1+w_\de}\delta_\de+\Psi\right) +\left(3c_{s\de}^2-1\right)v_\de
	\,.\\
\end{split}
\end{align}
and for the metric potential
\begin{align}\label{GTequations3n}
\begin{split}
\Psi_x+\Psi\left(1+\frac{k^2}{3\mathcal{H}^2}\right) &=-\frac{1}{2} \delta
\,,\\
\Psi_x+\Psi &=-\frac{3}{2}\mathcal{H}v\left(1+w\right)
\,,\\
\Psi_{xx}+\left[3-\frac{1}{2}\left(1+3w\right)\right]\Psi_x-3w\Psi &=\frac{3}{2}\frac{\delta p}{\rho} 
\,.\\
\end{split}
\end{align}
%

%
%

\subsection{The speed of sound of DE}
\label{speedsound_de}

So far, in this work we have described all the individual matter components as perfect fluids with a barotropic equations of state $p_A(\rho_A)$. Since a barotropic fluid is adiabatic, its effective and adiabatic squared speeds of sound are the same (cf. Eq.~\eqref{newvar0}). While for radiation and matter such a representation works well, for fluids with negative EoS, in particular for fluids playing the role of DE, there might be some problems if the squared speed of sound becomes negative, as this would lead to instabilities at the perturbative level. As a matter of extra-clarification we discuss in the Appendix~\ref{AppB} how instabilities at the linear level in perturbations arise in fluids with a negative adiabatic squared speed of sound. It is therefore necessary to take into account additional considerations for the DE fluid. To avoid this problem we note that the EoS presented in the previous section are effective descriptions of some  unknown fundamental field. As such, the barotropic nature of the models at the background level is not necessarily inherited by the cosmological perturbations.
 Bearing this in mind, we fix the effective squared speed of sound of DE, as defined in \eqref{newvar0}, to unity, i.e., in Eqs.~\eqref{seteq1each} we set $c_{s\de}^2=1$, while $c_{a\de}^2$ is given by Eq.~\eqref{newvar0}. 
We next show the expressions for the adiabatic speed of sound corresponding to the the models (i), (ii) and (iii), respectively.
\begin{align}\label{ca2express}
\begin{split}
\textrm{(i)}\qquad&
c_{a}^2=\BRparamA 
\,, 
\\
\textrm{(ii)}\qquad&
c_{a}^2=-\left(1+\frac{1}{2}\frac{\mathcal{B}}{\rho_{\textrm{d}}^{1/2}}\right)
\,,
\\
\textrm{(iii)}\qquad&
c_{a}^2=-1 \,
\end{split}
\end{align}

 This strategy can be encountered in several works in the literature \cite{Bean:2003fb,Valiviita:2008iv,Li:2013bya} and in cosmological codes  such as CAMB \cite{CAMB} and CLASS \cite{Lesgourgues:2011re}  in particular when interpreting the DE fluid as Quintessence. Here, we would like to point out that our choice of $c_{s\de}^2=1$ is purely phenomenological rather than deduced from a realistic theoretical grounded model. Nevertheless, as discussed in the Appendix~\ref{Mapping to a phantom scalar field}, this choice for $c_{s\de}^2$ extends to first order in perturbations the possibility of mapping the phantom DE fluid to a phantom scalar field.

%
%
 
\subsection{Initial conditions}
\label{initial_conditions}

Once the system of equations of the perturbed quantities is defined, we need to impose proper initial conditions in order to compute the cosmological evolution of the perturbations (cf. for example \cite{Ma:1995ey,Ballesteros:2010ks} for a detailed discussion on the initial conditions on DE models). For this goal, we will take into account the following considerations. First, we assume that at an initial moment, $z\sim 10^{6} $ (which roughly corresponds to $x_\ini\sim-14$), the Universe is completely dominated by radiation, so that all relevant quantities of the total matter fluid are those of a perfect fluid with $p=\rho/3$. Secondly, we note that at such moment the wave-length of all the relevant modes is small when compared with the comoving Hubble parameter ($k\ll\mathcal{H}$), i.e. they are outside the horizon. With these two approximations, we can combine the first and third equations of~\eqref{GTequations3n} and obtain a closed evolution equation for the metric potential in the asymptotic past \cite{AmendolaTsujikawa}
\begin{align}\label{PhieqRad}
	\Psi_{xx}+3\Psi_x \approx 0
	\,.
\end{align}
The dominant solution of this equation is a constant solution $\Psi_\ini=\Psi(x_\ini)$.
\footnote{For the rest of this section we will denote by $X_\ini$ the value of a quantity $X$ evaluated at $x=x_\ini$.
} Applying this result to the set of equations (\ref{GTequations3n}), we find that initially
\begin{align}
	\label{Phieq0}
	\begin{split}
		\Psi_\ini&\approx-\frac{1}{2} \delta_\ini
		\,,\\
		\Psi_\ini&\approx-2\mathcal{H}_\ini v_\ini
		\,.\\
	\end{split}
\end{align}

Assuming initial adiabatic conditions, we can relate the initial values of the individual fluid perturbed variables to the total perturbation in Eq.~\eqref{Phieq0} through \cite{AmendolaTsujikawa,Ma:1995ey,Ballesteros:2010ks}
\begin{align}
	\label{adiab}
	\frac{\delta_{\rad}}{1+w_{\rad}}=\frac{\delta_{\mat}}{1+w_{\mat}}=\frac{\delta_\de}{1+w_\de} = \frac{\delta}{1+w}
	\,.
\end{align}
This allows us to write the initial values of $\delta_A$ in terms of $\delta_\ini$ as
\begin{align}
	\label{initcond1}
	\frac{3}{4}\delta_{\rad,\ini}
	=
	\delta_{\mat,\ini}
	=
	\frac{\delta_{\de,\ini}}{1+w_{\de,\ini}}\approx\frac{3}{4}\delta_\ini
	\,,
\end{align}
By imposing the adiabatic condition \eqref{adiab}  between two fluids $A$ and $B$ on the comoving gauge, we obtain
\begin{align}
	\frac{\delta_{A,\ini} - 3\mathcal{H}_{\ini} \left(1+w_{A,\ini} \right) v_{A,\ini}}{1+w_{A,\ini}}
	= 
	\frac{ \delta_{B,\ini} - 3\mathcal{H}_\ini \left(1+w_{B,\ini} \right) v_{B,\ini} }{1+w_{B,\ini}}
	\,,
\end{align}
which then leads to the initial values of the peculiar velocities:
\begin{align}
	\label{velocityequal}
	v_{\rad,\ini} 
	= v_{\mat,\ini} 
	= v_{\de,\ini}
	\approx \frac{\delta_\ini}{4\mathcal{H}_\ini}
	\,.
\end{align}
We note that the conditions \eqref{initcond1} and \eqref{velocityequal} coincide with the ones presented in \cite{Ma:1995ey} in the absence of neutrinos.
Making use of the linearity of Eqs.~\eqref{seteq1each} and \eqref{GTequations3n}, we can first compute the evolution of the perturbation quantities using the initial conditions \eqref{Phieq0}, \eqref{initcond1} and \eqref{velocityequal} for $\Psi_\ini=1$ (which implies $\delta_\ini = -2$) and then multiply all the solutions obtained by the physical value of $\delta_\textrm{phys}(k)$, which we will take from the Planck observational fit to single field inflation \cite{Ade:2015xua}:
\begin{align}\label{cordelta}
	\delta_\textrm{phys}(k)=\frac{8\pi}{3} \sqrt{2A_{s}}\left(\frac{k}{k_{\textrm{pivot}}}\right)^{\frac{n_s-1}{2}}k^{-\frac{3}{2}}
	\,.
\end{align}
Here, $A_s$ and $n_s$ are defined as the amplitude and spectral index of the primordial inflationary power spectrum corresponding to a previously selected pivot scale $k_{\textrm{pivot}}$. In this work, we will use the values $k_{\textrm{pivot}}=0.05 \ \textrm{Mpc}^{-1}$, $A_s=2.143\times10^{-9}$, and $n_s=0.9681$ in accordance with Planck observational data \cite{Ade:2015xua}.%

\subsection{Matter power spectrum and the growth rate}

The matter power spectrum describes how galaxies are distributed along the Universe and provides us with a method to compare theoretical predictions with the observational data. In this work, we will compute the linear matter power spectrum for each model studied in Sec.~\ref{review} and we will try to detect deviations from the predictions of $\Lambda$CDM. The whole framework presented in the previous sections provide us the necessary tools to obtain the aimed results. Notice, however, that the correct definition of the matter power spectrum uses the fractional energy density perturbation in the comoving gauge \cite{Wands:2009ex,Bruni:2011ta}, while the analysis carried out in this work has been done in the Newtonian gauge. Using the variables employed in the previous sections, we can resolve this gauge difference by expressing the matter power spectrum as
\begin{align}\label{mpsP}
	P_{\hat{\delta}_{\mat}} = \left\vert\delta_{\mat}^{(com)}\right\vert^2 
	=\left\vert\delta_{\mat} - 3\mathcal{H}v_{\mat}\right\vert^2
	\,.
\end{align}

Another method we will use to constrain the models presented in Section \ref{review} is based in computing the growth rate of the matter perturbations for the different models. By definition, the growth rate of the matter perturbations is given by the formula \cite{Balcerzak:2012ae}
\begin{align}\label{f_definition}
	f\equiv\frac{d \left(\ln\delta_{\mat}\right)}{d \left(\ln a\right)}
	\,.
\end{align}
For DM-DE models that closely mimic $\Lambda$CDM, it was found that the growth rate at late-time can be approximated reasonably well by the formula \cite{Wang:1998gt,Linder:2005in,Linder:2007hg}
\begin{align}\label{f_approx}
	f\simeq\Omega_{\mat}^{\gamma}
	\,,
\end{align}
where $\gamma\simeq0.55$ for $\Lambda$CDM. The next to leading order of (\ref{f_approx}) can be found in \cite{wikiesa}.

In this work, instead of using any approximated parametrisation, we opt to calculate the evolution of the growth rate $f$ for each DE model using the full Eqs.~\eqref{seteq1each}-\eqref{GTequations3n} and compare the results with observations.
We note, however, that in most cases, the observational data refers not to the growth rate $f$ directly, but to the combination $f\sigma_8$, where $\sigma_8$ is the root mean square mass fluctuation amplitude in spheres of size $8\textrm{h$^{-1}$Mpc}$ which is used to normalise the matter power spectrum.
This combination has the advantage that it avoids the degeneracy in the parameter space regarding $\sigma_8$ and the linear bias, $b$, between the perturbations of dark matter and density of galaxies \cite{Song:2008qt}.
We calculate the temporal evolution of $\sigma_8$ by the formula%
\footnote{We thank V. Salzano for clarifying this point to us.}
\cite{Wang:2010gq}
\begin{align}\label{sigma8}
	\sigma_8\left(z,\,k_{\sigma_8}\right) = \sigma_8\left(0,\,k_{\sigma_8}\right) \frac{\delta_{\mat}\left(z,\,k_{\sigma_8}\right) }{\delta_{\mat}\left(0,\,k_{\sigma_8}\right)}
	\,.
\end{align}
where $k_{\sigma_8}=0.125$ hMpc$^{-1}$ is the wave-length of the mode corresponding to distances of $8\textrm{h$^{-1}$Mpc}$.

For each of the DE models considered in this work, we will calculate the evolution of $f\sigma_8$ using the numerical solutions of Eqs.~\eqref{seteq1each}-\eqref{GTequations3n} and the relations \eqref{f_definition} and \eqref{sigma8}.
 For all the models we use $\sigma_8(0,k_{\sigma_8})=0.820$ \cite{wikiesa} as the present day value of $\sigma_8$.
We compare the results obtained with the available observational data \cite{Hudson:2012gt,Beutler:2012px,Howlett:2014opa,Percival:2004fs,Song:2008qt,Blake:2011rj,Samushia:2011cs,Tojeiro:2012rp,Gil-Marin:2015sqa,Tegmark:2006az,Blake:2012pj,Chuang:2013wga,Guzzo:2008ac,delaTorre:2013rpa,Okada:2015vfa,Satpathy:2016tct} to check whether the predictions of the models are within the observational constraints. Since at the background level these models are very similar to $\Lambda$CDM till the present time, we expect that the deviations from $\Lambda$CDM in the evolution of the growth rate to be small.

%
%

\section{Results}
\label{results}

In this section, we present and discuss the results obtained for the evolution of the cosmological perturbations in the three models discussed in Sec~\ref{review} that contain distinct future cosmological abrupt events:
\begin{enumerate}[(i)]
\item a Big Rip singularity (Sec.~\ref{BR}),
\item a Little Rip event (Sec.~\ref{LR}) 
\item a Little Sibling of the Big Rip event (Sec.~\ref{LSBR}).
\end{enumerate}

For each of these models, the evolution of the matter perturbations $\delta_{\mat}$, $v_{\mat}$, $\delta_{\rad}$, $v_{\rad}$, $\delta_\de$, and $v_\de$ was obtained by numerically integrating the set of Eqs. \eqref{seteq1each} after substituting $\Psi$ and $\Psi_{x}$ given in \eqref{GTequations3n}. After carrying this numeric integration, the gravitational potential $\Psi$ and its derivative $\Psi_{x}$ can be obtained from the first two equations in \eqref{GTequations3n}.
The integration was performed since an initial moment deep inside the radiation epoch ($z\sim 10^{6} $), when all the relevant modes are outside the horizon, and till a point in the distant future ($z\sim -0.99$).
At the initial moment, the values of the variables were fixed according to Eqs. \eqref{initcond1}, \eqref{velocityequal} and \eqref{cordelta}. In addition, for each model this integration was repeated for 200 different modes with wave-numbers ranging from $k_\textrm{min}\sim3.3\times10^{-4}\textrm{h Mpc}^{-1}$, which corresponds to the mode that is exiting the Hubble horizon at the present time, to a $k_\textrm{max}\sim3.0\times10^{-1}\textrm{h Mpc}^{-1}$. Notice that for $k\gtrsim k_\textrm{max}$ the validity of the linear perturbation theory breaks down as non-linear effects start to become dominant in the evolution of the perturbations.
In fact, $k_\textrm{max}$ should be at most $2.0\times10^{-1}\textrm{h Mpc}^{-1}$. On the plots we included the higher value $k_\textrm{max}\sim3.0\times10^{-1}\textrm{h Mpc}^{-1}$ to amplify visually the effect and evolution on the largest modes. 
 As we mention in Sec. \ref{review}, the cosmological parameters $\Omega_{\textrm{m0}}$ and $H_0$ for all the three models studied were taken from the recent Planck mission \cite{Ade:2015xua}. 
While the value of $w_d$ for the model (i) was fixed according to the Planck data for wCDM model \cite{wikiesa}, the parameter $\LRparamA$ of model (ii) and the parameter $\LSBRparamA$ of model (iii) were fixed so that in all models the variable $\delta_\de$ has the same value at the initial moment. 
\footnote{
This implies that initially all the models have the same value for the DE EoS.
}
As such, the results of this section should be viewed as a first step in obtaining a description of the evolution of the cosmological perturbations in phantom DE models. A more realistic picture of the imprints of each model will be explored in a future work where a fit of the parameters of the models will be performed using the available cosmological data.

\begin{figure}[ht!]
 \includegraphics[width=1\textwidth]{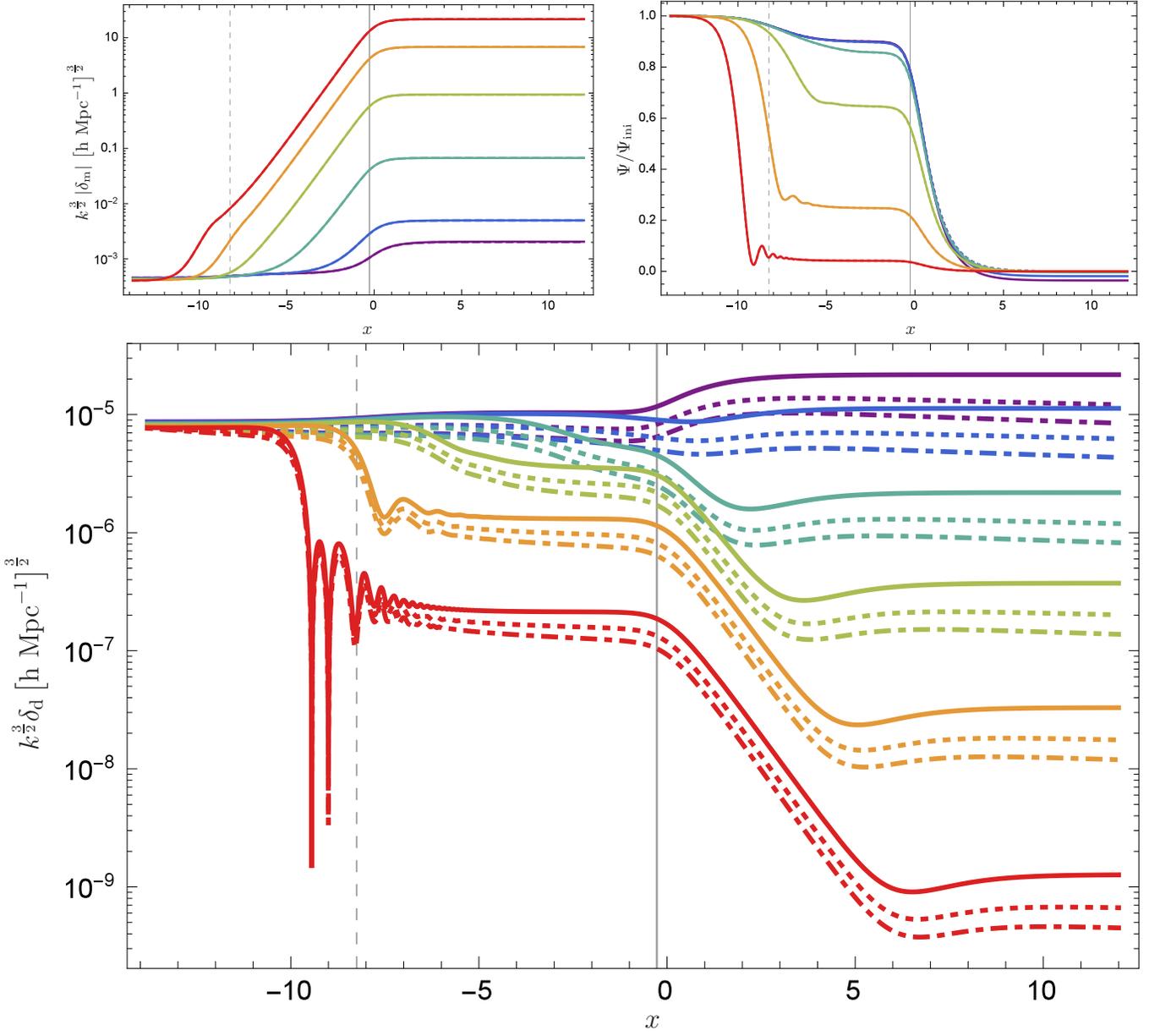}
\caption{\label{perturbationsgraphs}%
The top panels of this figure show the evolution of (top-left) the matter perturbation, $\delta_\mat$, and (top-right) the gravitational potential, $\Psi$, the latter being normalised to its initial value $\Psi_\ini$, for different modes $k$. All the three models considered present an almost identical behaviour that makes them indistinguishable from $\Lambda$CDM (such is the case that although we have plotted the results of $k^{3/2}\left\vert\delta_{\textrm{m}}\right\vert$ and $\Psi/\Psi_\ini$ for $\Lambda$CDM, we cannot distinguish those results from the others). The bottom panel shows the evolution of the perturbation of DE, $\delta_\de$, for the same modes. Here, the differences between the three models become more noticeable, in particular in the amplitude of the perturbations.
In all panels the solid lines correspond to the model (i), the dotted lines to the model (ii) and the dot-dashed lines to the model (iii).
 Each colour represents a different mode: $k=3.33\times10^{-4}\textrm{h Mpc}^{-1}$ (purple), $k=7.93\times10^{-4}\textrm{h Mpc}^{-1}$ (dark blue), $k=3.50\times10^{-3}\textrm{h Mpc}^{-1}$ (light blue), $k=1.54\times10^{-2}\textrm{h Mpc}^{-1}$ (green), $k=6.80\times10^{-2}\textrm{h Mpc}^{-1}$ (orange), $k=0.30 \textrm{h Mpc}^{-1}$ (red). All perturbations are represented versus $x=\log(a/a_0)$ which varies from values well inside the radiation era ($x=-13.81$) till the far future ($x=12$). The value $x=0$ corresponds to the present time. The dashed vertical line corresponds to the radiation-matter equality while the solid vertical line represents the equality between DE and matter.
 }%
\end{figure}

\begin{figure}[ht!]
 \includegraphics[width=1\textwidth]{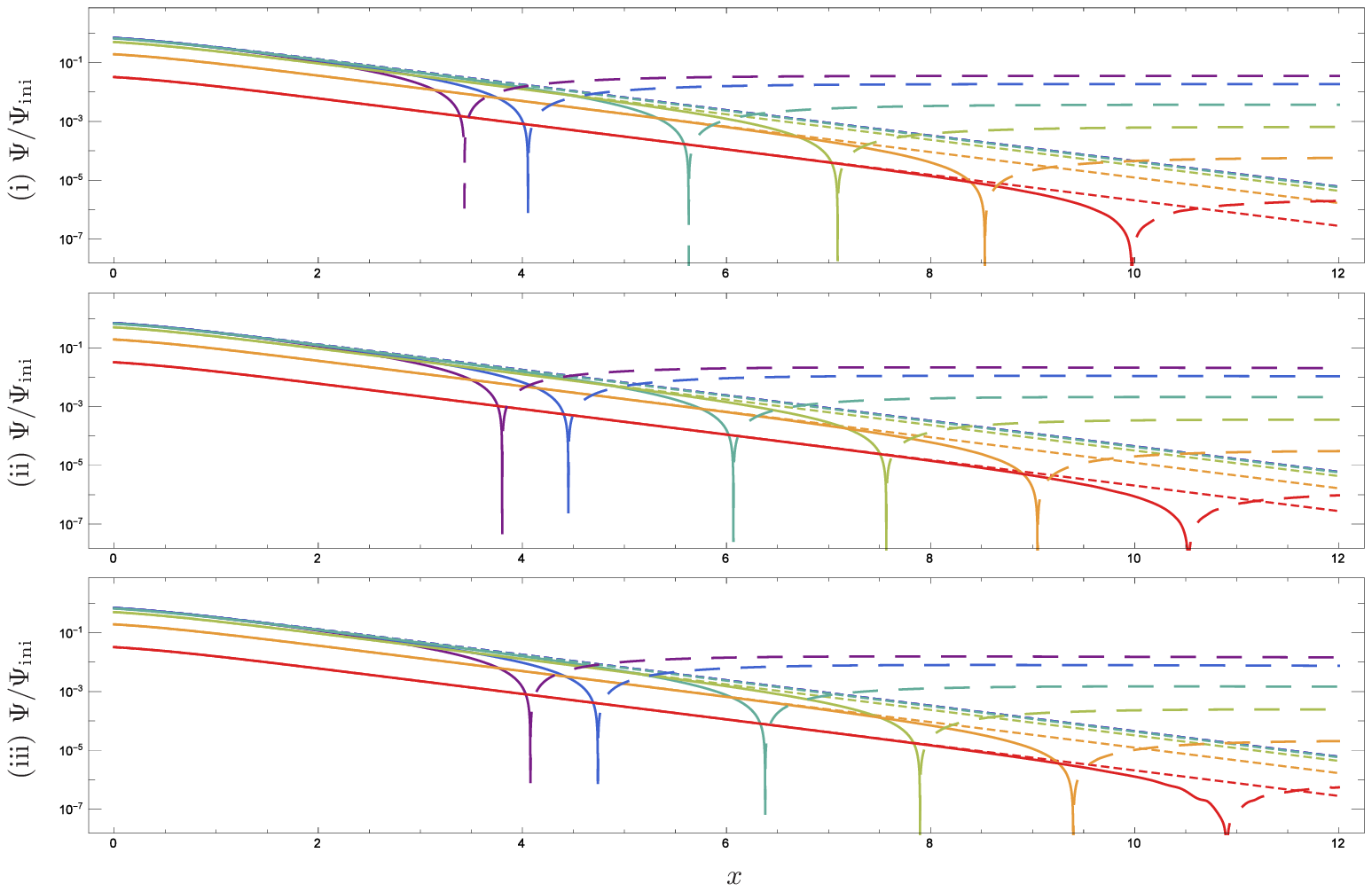}
\caption{\label{phiphiphi}%
This Figure shows the evolution of the gravitational potential $\Psi$ for the model (i) on the top-panel, for the model (ii) on the middle panel, and for the model (iii) on the bottom panel for the same modes $k$ shown in Fig.~\ref{perturbationsgraphs}. Positive (negative) values of $\Psi$ are indicated with solid (dashed) lines. On all panels the corresponding behaviour in $\Lambda$CDM model is indicated by a dotted line. While within a $\Lambda$CDM model the gravitational potential decays exponentially with positive values of $x$ until the asymptotic future, in the phantom DE models (i), (ii) and (iii) $\Psi$ approaches constant negative negative value in the future. The change in sign of $\Psi$ is scale and model dependent - for a given model it happens first for the larger scales (smaller $k$) and for the same mode it happens first for the model (i), then for the model (ii) and then for the model (iii).
 }
\end{figure}

In Fig.~\ref{perturbationsgraphs}, we illustrate the evolution of the cosmological perturbations, since the initial moment and till a point in the future evolution of the Universe. The two top panels show the evolution (top-left) of the fractional energy densities of DM, $\delta_{\mat}$, and (top-right) of the gravitational potential, $\Psi$, which has been normalised with respect to its initial value, $\Psi_\ini$, while the bottom panel shows the evolution of the fractional energy densities of DE, $\delta_{\de}$.
In each panel we identify the results of the model (i) using solid lines, the results of the model (ii) using dotted lines and the results of the model (iii) using dot-dashed lines. 
For each of these quantities, we plot the results for 6 different wave-numbers: $k=3.33\times10^{-4}\textrm{h Mpc}^{-1}$ (purple), $k=7.93\times10^{-4}\textrm{h Mpc}^{-1}$ (dark blue), $k=3.50\times10^{-3}\textrm{h Mpc}^{-1}$ (light blue), $k=1.54\times10^{-2}\textrm{h Mpc}^{-1}$ (green), $k=6.80\times10^{-2}\textrm{h Mpc}^{-1}$ (orange), and $k=0.30 \textrm{h Mpc}^{-1}$ (red).
In terms of evolution, we can distinguish three different behaviours, according to the range of the wave-numbers:
\begin{itemize}
\item large k: 0.30 hMpc$^{-1}$ (red) and $ 6.80 \times10^{-2}$hMpc$^{-1}$ (orange). 
\item medium k: $1.54\times10^{-2}$ hMpc$^{-1}$ (green) and $3.50 \times10^{-3}$hMpc$^{-1}$ (light blue).
\item small k: $7.93\times10^{-3}$ hMpc$^{-1}$ (dark blue) and $3.33\times10^{-4}$ hMpc$^{-1}$ (purple). 
\end{itemize}

The top-left panel of Fig.~\ref{perturbationsgraphs} shows that the evolution of the matter perturbations for the different models and for $\Lambda$CDM presents an almost identical behaviour. During the initial radiation dominated epoch, each individual mode remains constant until it enters the Hubble horizon. After this point, the gravitational collapse leads to the growth of $\delta_{\mat}$, which becomes exponential in $x$ during the matter era. Once the DE starts to become dominant, the growth of the matter perturbations slows down as $\delta_{\mat}$ seems to converge to a constant value in the asymptotic future.
Notice that since the modes with larger $k$ enter the horizon at earlier time, they correspond to the curves with higher values of $\delta_{\mat}$ in Fig.~\ref{perturbationsgraphs}.


The top-right panel of Fig.~\ref{perturbationsgraphs} presents the evolution of the gravitational potential $\Psi$. 
As in the case of $\delta_\mat$, the overlap between the three models studied and $\Lambda$CDM is almost perfect and any differences until the present time are virtually undetectable. 
For all the ranges of $k$  considered the perturbations start with a constant value at the radiation dominated epoch, reach a second plateau during  the matter dominated era and start decaying when DE starts to dominate. Nevertheless, we can identify some qualitative differences in the evolution of $\Psi$ until the present time, depending on the range of $k$:
\begin{itemize}
\item large $k$:  These modes enter the Horizon during the radiation dominated era. Around the time of horizon crossing they start to decay and can present oscillations before the moment of radiation-matter equality. Then, during the matter dominated era the oscillations are suppressed and the perturbations reach a constant value till DE gains importance.

\item medium $k$: The gravitational potential remains constant during the radiation dominated epoch, decays around the radiation-matter equality and reaches a second plateau in the matter dominated era. The decay observed in the transition between the two epochs is scale dependent and affects mostly the modes with higher wave-number. 

\item small $k$: As in the previous case, the modes corresponding to the smallest wave-numbers show a constant behaviour during the radiation dominated and the matter dominated epochs and a decay around the radiation-matter equality. Here, however, the decay is scale independent and for all the modes the amplitude of $\Psi/\Psi_\ini$ during the matter era is $9/10$ of its initial value, as follows from theoretical prediction in the limit $k\rightarrow0$ \cite{AmendolaTsujikawa}.
\end{itemize}

Around the matter-DE equality,  we find that for all modes the amplitude of the gravitational potential starts to decay rapidly. However, after some 5 e-folds of expansion into the future, the top-right panel of Fig.~\ref{perturbationsgraphs} seems to indicate that the value of the modes of the gravitational potential stabilises at a negative value. In order to have a clearer picture of this behaviour, we plot in Fig.~\ref{phiphiphi} the evolution of $|\Psi/\Psi_\ini|$ in logarithmic scale from $x=0$ to $x=12$ for:  top panel - model (i); middle panel - model (ii); and bottom panel - model (iii). Here, we see that after an initial period of exponential decay, eventually the value of $\Psi$ changes sign  and then evolves towards a negative constant (negative values of $\Psi$ are indicated a dashed line). This behaviour is in clear contrast with the evolution in the $\Lambda$CDM model, indicated by dotted lines, where we see that the exponentially decay with respect to $x$ continues asymptotically. Although not depicted here, it was found that for quintessence models with constant $w>-1$ the gravitational potential also evolves towards a constant but with a positive asymptotic value. Therefore, this change in sign of the gravitational potential appears as a clear indicator of a phantom evolution. Notice however that this only happens in the far future and cannot be observed at the present time. From Fig.~\ref{phiphiphi} it can also be seen that this effect is scale dependent, for the same model the change in sign happens first for the larger scales and only later for the smaller ones, and model dependent, for the same value of $k$ the change in sign happens first for the model (i), then for the model (ii) and finally for the model (iii). 

In order to understand the effects described above, we note that once the DE dominated era begins and the modes start to exit the horizon again, the gravitational potential is initially sourced by the matter perturbations and DE perturbations. Therefore, the first equation in~\eqref{GTequations3n} can be written as
\begin{align}
	\label{asimptotic_evol}
	\Psi_x+\Psi\left(1+\frac{k^2}{3\mathcal{H}^2}\right) &=-\frac{1}{2} \left(\Omega_\mat\delta_\mat + \Omega_\de\delta_\de\right)
	\,.
\end{align}
If the rhs of Eq.~\eqref{asimptotic_evol} evolves asymptotically to a positive constant, then the potential stabilises at a positive value. On the other hand, if the potential is to cross the $\Psi=0$ value and stabilise at a negative value than the the rhs of Eq.~\eqref{asimptotic_evol} needs to evolve towards a negative constant in the far future.
We now recall that $\Omega_\mat$ is decreasing exponentially with $x$ while $\Omega_\de$ is approaching unity, and that due to the adiabatic conditions \eqref{initcond1} imposed at the beginning of the integration, $\delta_\mat$ is negative-valued while for a phantom fluid $\delta_\de$ is positive-valued. Thus, the changing of sign of the gravitational potential can happen in a phantom DE model after the equality 
\begin{align}
	\label{matter_de_equality}
	|\Omega_\mat\delta_\mat| = |\Omega_\de\delta_\de|
	\,,
\end{align}
is reached. From Fig.~\ref{perturbationsgraphs} we observe that when $\delta_\mat$ and $\delta_\de$ become constant after matter-DE equality, $\left\vert\delta_\mat\right\vert$ is larger for the modes with larger $k$ while $\delta_\de$ larger for the modes with smaller $k$. Therefore, the modes corresponding to smaller scales are the ones that see the change in the sign of $\Psi$ first. On the other hand, for the model (i) the ratio $\Omega_\mat/\Omega_\de$ decays faster than for the models (ii) and (iii), while decaying faster for the model (ii) than for the model (iii). Therefore, for the same value of $k$ the equality \eqref{matter_de_equality} is reached first for model (i), then for model (ii) and finally for model (iii), as seen in Fig.~\ref{phiphiphi}.

The bottom panel of Fig.~\ref{perturbationsgraphs}, we present the evolution of the fractional energy density of DE for different wave-numbers. In contrast with the perturbations of DM, here we observe some differences between the three models.  First, we note that while the initial value of $\delta_{\de,\ini}$ is the same for all the models, at the present time for the models (ii) and (iii) $\delta_{\de}$ appears to be systematically suppressed on all the scales with regards to the  model (i). 
This suppression can be understood by the evolution of  $\delta_{\de}$ deep inside the radiation dominated and the matter dominated epochs. For model (i) we observe in Fig.~\ref{perturbationsgraphs} that all the modes present a constant plateau in these periods. in contrast, in the models (ii) and (iii) we find that similar plateaus  exist in these periods but with a negative tilt, meaning that the amplitude of $\delta_\de$ is continuously decreasing until the present time. This effect seems to be tied to how strong is the variation of $w_\de$ during these periods since the model (iii), which is the one that more rapidly converges to $w_\de\approx-1$ is the one that sees a stronger suppression of the DE perturbations.

Similarly to what happens with the DM perturbations, we can characterise the qualitative evolution of $\delta_\de$ in terms of the range of the wave-number of the mode:
\begin{itemize}
\item large $k$: These modes are the first to enter the horizon during the radiation epoch.  While initially their amplitude decreases slowly, after the horizon crossing they suffer a fast decay and we observe a damped oscillatory behaviour till the radiation-matter equality. At this point the modes present a second plateau and that lasts until the end of the matter era. Once DE starts to dominate they start to decay rapidly once more.
\item medium $k$: In this range, the modes present a behaviour of transition. While they enter the horizon only in the matter era, and therefore present no early oscillations, we still observe the existence of a plateau during the matter dominated era. The length of the plateau depends on the horizon crossing time - for the modes with smaller $k$ the amplitude does not have time to stabilise before the DE dominated epoch. At this point, the amplitude starts to decay but, much like what happens for modes in the small $k$ range, it stabilises once the accelerated expansion shrinks the Hubble radius and leads the modes to exit the horizon.
\item small $k$: These modes are outside of the horizon for most of their evolution and therefore present little to no variation in amplitude till close to the present time, when they enter the horizon. At this point we observe a difference in behaviour, with the modes with smallest wave-numbers being amplified during the initial stages of the DE dominated epoch, while the largest modes in this range present a slight decay. Once the DE fluid completely dominates, however, the modes exit the horizon and their amplitude stabilises once more.
\end{itemize}
Despite all the differences in behaviour and amplitude between distinct modes, we always find that the amplitude of DE perturbations is extremely small when compared to DM. This validates the usual assumption of a DE smooth fluid.

\begin{figure}[ht]
 \includegraphics[width=\textwidth]{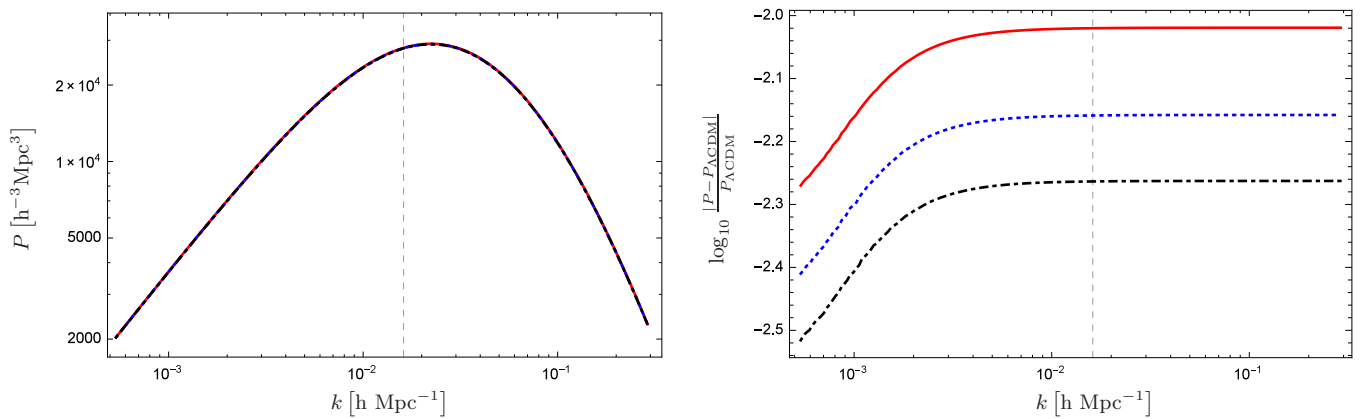}
\caption{\label{diflambda}%
The lhs panel shows the almost perfect superposition of the matter power spectra, $P_{\delta_\mat}$, of the models (i), (ii) and (iii) with the matter power spectrum of $\Lambda$CDM.  For clarity of the plot, we did not plot the case of $\Lambda$CDM as it overlaps perfectly with the other three curves.
The rhs panel shows the relative deviation of each model in comparison with $\Lambda$CDM. In both panels, the red  solid curve corresponds to the model (i), the blue  dotted curve corresponds to the model (ii) and the black  dot-dashed curve corresponds to the model (iii).
 The dashed vertical line denotes the mode that crossed the horizon at the moment of the radiation-matter equality. All the models present a small enhancement ($\lesssim1\%$) of the amplitude of $P_{\delta_\mat}$. This effect is increasingly suppressed for large scales but seems to become scale invariant for modes that are already inside the horizon during the radiation-matter equality. 
The model that induces a BR singularity shows the highest enhancement in the power spectrum while the model that induces a LSBR abrupt event presents the smallest deviation respect to the $\Lambda$CDM model.
}
\end{figure}

On the left hand side (lhs) of Fig.~\ref{diflambda}, we compare theoretical matter power spectrum, $P_{\delta_{\mat}}$, for each of the models considered in this work with the one predicted by $\Lambda$CDM. For all the cases, we find an almost perfect superposition of the spectra. This reflects the close resemblance in terms of evolution of all the models up to the present time and suggests that observables like the matter power spectrum may not be able to distinguish dark energy models that do not differ significantly from $\Lambda$CDM till today.
In each panel we identify the results of model (i) using solid lines, the results of model (ii) using dotted lines and the results of model (iii) using dot-dashed lines.

In order to be able to make a comparison between the results of the models (i), (ii) and (iii) we have proceeded to plot on the rhs panel of Fig.~\ref{diflambda} the relative difference in the magnitude of the matter power spectrum with respect to the $\Lambda$CDM model. 
For all the models we find a small enhancement ($\lesssim1\%$) in the amplitude of the matter power spectrum that is practically constant for modes that are outside of the horizon at the moment of the radiation-matter equality. At the large-scale end of the spectrum, we find that this enhancement becomes increasingly small. This effect seems to be related to how much the model in question deviates from $\Lambda$CDM until the present time: model (i) which is the one that deviates the most from $\Lambda$CDM is the one that sees a stronger enhancement, while model (iii) is the one that more closely resembles $\Lambda$CDM and sees the faintest effect. These results are in conformity with Fig.~6 of Ref.~\cite{Caldwell:1999ew} where it is shown that in $w$CDM the suppression of the growth of the matter perturbations becomes smaller as $(1+w_\de)$ becomes more and more negative.

 Finally, we present in Fig.~\ref{growth} the evolution of $f\sigma_8$ (lhs panel) and the relative deviation of $f\sigma_8$ with respect to $\Lambda$CDM (rhs panel) for the three models studied and for a redshift within the range $z\in(0,\,1.4)$. 
In each panel we identify the results of the model (i) using solid lines, the results of the model (ii) using dotted lines and the results of the model (iii) using dot-dashed lines. 
We find that all models are within the error bars for almost all the points (cf. Table \ref{tablegrowth}). Nevertheless, there seems to be some tension between the theoretical predictions and the observational data, as the $f\sigma_8$ curves are systematically above most of the data points for redshifts up to $z\sim0.8$. This tension between the theoretical predictions based on CMB data -- higher values of $\Omega_\mat$ and $\sigma_8$ -- and the local redshift distortion measurements -- lower values of $\Omega_\mat$ and $\sigma_8$ -- is already found in $\Lambda$CDM and is not a special feature of the models studied in this work. A more detailed discussion on this topic can be found in \cite{Macaulay:2013swa,Battye:2014qga}.}

 On the rhs panel of Fig.~\ref{growth}, which presents the relative deviation of each model with regards to $\Lambda$CDM, we find the same tendency as in the case of the matter power spectrum in Fig.~\ref{diflambda}: there is an enhancement of $f\sigma_8$ for all models ($\lesssim0.4\%$) that seems to be more intense the more the model deviates from $\Lambda$CDM, i.e., the effects are more intense for the model (i), followed by the model (ii) and finally for the model (iii). These results are in conformity with the increase in the growth of the matter perturbations in a phantom scenario shown in Fig.~6 of Ref.~\cite{Caldwell:1999ew}.
After the matter-DE equality at $z\sim0.3-0.4$, this effect starts to vanish rapidly, with the deviations from $\Lambda$CDM being $\lesssim0.08\%$ around the present time. This behaviour seems to be associated to the fact that in all models $\delta_\mat$ becomes constant at late-time, when DE completely dominates the energy budget of the Universe.

By simply looking at the plot on the lhs of Fig.~\ref{growth}, we can clearly see that the three considered models fit pretty well the observations. We can try to understand which of these three models fits better the observational data. For this we would need to make a fitting of the background models which is far beyond the current work. Therefore, we will simply ``extrapolate'' the definition of the reduced $\chi^2$:

\begin{align}
	\chi^{2}=\frac{1}{N- N_{\textrm{fp}}}\sum_{i}^{N}\frac{\left[f_{obs}\left(z_i\right)-f_{th}\left(z_i\right)\right]^{2}}{\sigma_{i}^{2}}
	\,.
\end{align}
Here, $f_{obs}\left(z_i\right)$ and $f_{th}\left(z_i\right)$ are, respectively, the observational and theoretical growth rates at redshift $z_i$, while $\sigma_{i}$ is the corresponding error for each measurement and $N$ the total number of observations while $N_{\textrm{fp}}$ is the number of fitted parameters. We will work out these numbers for  $N_{\textrm{fp}}=3$ as our model has, or  $N_{\textrm{fp}}=0$ as we did not fit any of them. Likewise, we will do it for the $\Lambda$CDM model. Our results are shown in table \ref{chitab}. These results seem to suggest that although all the models considered in this work provide a good fit to the observational data, this fit tends to become worse as the background evolution deviates more from $\Lambda$CDM.

\begin{figure}[t]\label{growthrate1}
 \includegraphics[width=\textwidth]{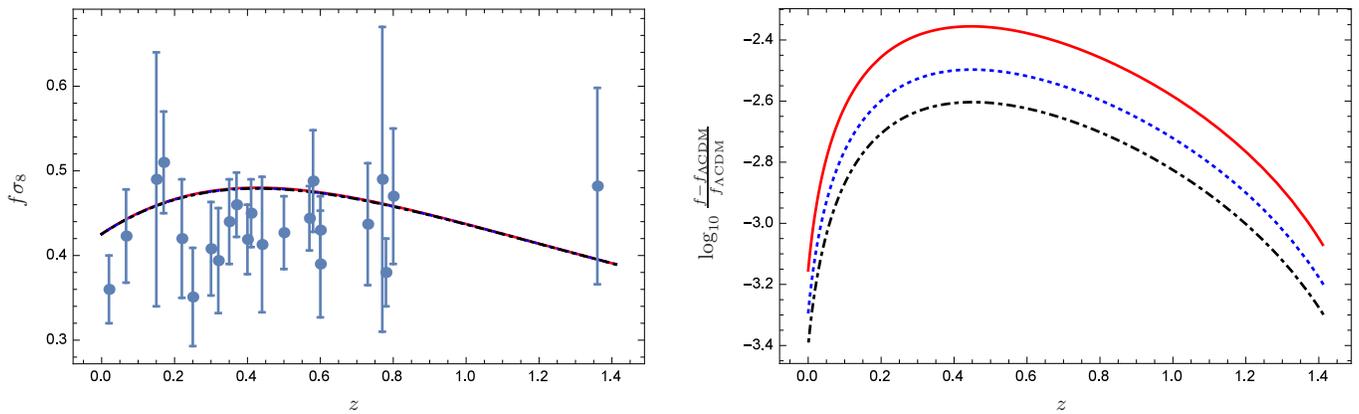}
\caption{\label{growth}%
The lhs panel of this figure shows the evolution of $f\sigma_8$ in terms of the redshift against the available data points indicated in Table~\ref{tablegrowth}. The red solid curve corresponds to the model (i), the blue dotted curve corresponds to the model (ii) and the black dot-dashed curve corresponds to the model (iii). For clarity of the plot, we did not plot the case of $\Lambda$CDM as it overlaps with the other three curves.
The rhs panel shows the relative deviation of $f\sigma_8$ with respect to $\Lambda$CDM. 
It can be shown that for the models there is a small enhancement  of $f\sigma_{8}$ ($\lesssim0.4\%$), which starts to decrease rapidly after the matter-DE equality ($z\sim 0.3-0.4$). 
The model that induces a BR singularity shows the highest enhancement in the power spectrum while the model that induces a LSBR abrupt event presents the smallest deviation respect to the $\Lambda$CDM model.
}
\end{figure}

\begin{table}[t]
\centering
\begin{tabularx}{.5\textwidth}{ C{1.4} C{.8}  C{.8} | C{1}  C{1}}
\toprule
	{\centering{\bf{Model}}}
	 &  \multicolumn{2}{c}{$N_{\textrm{fp}}$}  
	 &  \multicolumn{2}{c}{$\chi^{2}$}
  \\\midrule
\multicolumn{1}{c|}{$\Lambda$CDM}  &  $2$   &    $0$    &  1.094  &  1.010  \\
\multicolumn{1}{c|}{(i)} & $3$ & $0$ & 1.202  &  1.063 \\
\multicolumn{1}{c|}{(ii)} &  $3$ & $0$ & 1.185  &  1.048 \\
\multicolumn{1}{c|}{(iii)} &  $3$   & $0$ & 1.176  &  1.040 \\
\bottomrule
\end{tabularx}
\caption{\label{chitab}%
This table shows the reduced $\chi^2$ for different values of $N_{\textrm{fp}}$. The lhs (rhs) column within the $N_{\textrm{fp}}$ column corresponds to the lhs (rhs) column within the $\chi^2$ column.}
\end{table}

\begin{table}[ht!]
\centering
\begin{tabularx}{.85\textwidth}{ C{0.3} C{.7} C{2.6} C{0.4} }
\toprule
	{\centering{\bm{$z$}}}
	& \bm{$f\sigma_8$}
	& {\bf Survey}
	& {\bf Ref.}
\\\midrule
	$0.02\phantom{0}$
	& $0.36\phantom{0}\pm0.04\phantom{0}$
	& 
	& \cite{Hudson:2012gt}
 \\\midrule
	$0.067$
	& $0.423\pm0.055$
	& 6dF Galaxy Survey
	& \cite{Beutler:2012px}
 \\\midrule
	$0.15\phantom{0}$
	& $0.49\phantom{0}\pm0.15\phantom{0}$
	& SDSS DR7 MGS
	& \cite{Howlett:2014opa}
 \\\midrule
	$0.17\phantom{0}$
	& $0.51\phantom{0}\pm0.06\phantom{0}$
	& 2dF Galaxy Redshift Survey
	& \cite{Percival:2004fs,Song:2008qt}
 \\\midrule
	$0.22\phantom{0}$
	& $0.42\phantom{0}\pm0.07\phantom{0}$
	& WiggleZ Dark Energy Survey
	& \cite{Blake:2011rj}
 \\\midrule
	$0.25\phantom{0}$
	& $0.351\pm0.058$
	& SDSS II LRG
	& \cite{Samushia:2011cs}
 \\\midrule
	$0.3\phantom{00}$
	& $0.407\pm0.055$
	& SDSS I/II LRG + SDSS III BOSS CMASS
	& \cite{Tojeiro:2012rp}
 \\\midrule
	$0.32\phantom{0}$
	& $0.394\pm0.062$
	& SDSS III BOSS DR12 LOWZ
	& \cite{Gil-Marin:2015sqa}
 \\\midrule
	$0.35\phantom{0}$
	& $0.440\pm0.05\phantom{0}$
	& SDSS DR5 LRG
	& \cite{Song:2008qt,Tegmark:2006az}
 \\\midrule
	$0.37\phantom{0}$
	& $0.460\pm0.038$
	& SDSS II LRG
	& \cite{Samushia:2011cs}
 \\\midrule
	$0.38\phantom{0}$
	& $0.430\pm0.054$
	& SDSS III BOSS DR12
	& \cite{Satpathy:2016tct}
 \\\midrule
	$0.4\phantom{00}$
	& $0.419\pm0.041$
	& SDSS I/II LRG + SDSS III BOSS CMASS
	& \cite{Tojeiro:2012rp}
 \\\midrule
	$0.41\phantom{0}$
	& $0.45\phantom{0}\pm0.04\phantom{0}$
	& WiggleZ Dark Energy Survey
	& \cite{Blake:2011rj}
 \\\midrule
	$0.44\phantom{0}$
	& $0.413\pm0.080$
	& WiggleZ Dark Energy Survey + Alcock-Paczynski distortion
	& \cite{Blake:2012pj}
 \\\midrule
	$0.5\phantom{00}$
	& $0.427\pm0.043$
	& SDSS I/II LRG + SDSS III BOSS CMASS
	& \cite{Tojeiro:2012rp}
 \\\midrule
	$0.51\phantom{0}$
	& $0.452\pm0.057$
	& SDSS III BOSS DR12
	& \cite{Satpathy:2016tct}
 \\\midrule
	$0.57\phantom{0}$
	& $0.444\pm0.038$
	& SDSS III BOSS DR12 CMASS
	& \cite{Gil-Marin:2015sqa}
 \\\midrule
	$0.59\phantom{0}$
	& $0.488\pm0.06\phantom{0}$
	& SDSS III BOSS DR12 CMASS
	& \cite{Chuang:2013wga}
 \\\midrule
	$0.60\phantom{0}$
	& $0.43\phantom{0}\pm0.04\phantom{0}$
	& WiggleZ Dark Energy Survey
	& \cite{Blake:2011rj}
 \\\midrule
	$0.6\phantom{00}$
	& $0.433\pm0.067$
	& SDSS I/II LRG + SDSS III BOSS CMASS
	& \cite{Tojeiro:2012rp}
 \\\midrule
	$0.60\phantom{0}$
	& $0.390\pm0.063$
	& WiggleZ Dark Energy Survey + Alcock-Paczynski distortion
	& \cite{Blake:2012pj}
 \\\midrule
	$0.61\phantom{0}$
	& $0.457\pm0.052$
	& SDSS III BOSS DR12
	& \cite{Satpathy:2016tct}
 \\\midrule
	$0.73\phantom{0}$
	& $0.437\pm0.072$
	& WiggleZ Dark Energy Survey + Alcock-Paczynski distortion
	& \cite{Blake:2012pj}
 \\\midrule
	$0.77\phantom{0}$
	& $0.490\pm0.18\phantom{0}$
	& VIMOS-VLT Deep Survey
	& \cite{Guzzo:2008ac,Song:2008qt}
 \\\midrule
	$0.78\phantom{0}$
	& $0.38\phantom{0}\pm0.04\phantom{0}$
	& WiggleZ Dark Energy Survey
	& \cite{Blake:2011rj}
 \\\midrule
	$0.8\phantom{00}$
	& $0.470\pm0.08\phantom{0}$
	& VIMOS Public Extragalactic Redshift Survey
	& \cite{delaTorre:2013rpa}
 \\\midrule
	$1.36\phantom{0}$
	& $0.482\pm0.116$
	& FastSound
	& \cite{Okada:2015vfa}
 \\
\bottomrule
\end{tabularx}
\caption{\label{tablegrowth}%
This table shows the available observational data points for $f\sigma_8$ at different redshifts, which are plotted in Fig.~\ref{growth}. For each data point we present, in order, the value of the effective redshift, the value of $f\sigma_8$ and respective error, the corresponding survey, and the reference from which the values were taken.%
}
\end{table}

%
%

\section{Conclusions}
\label{conclusions}

In this work, we analyse the cosmological perturbations within the framework of GR, taking into full account the presence of DE at the perturbative level. The DE component is described by three different models, where each one of them behaves almost as $\Lambda$CDM model at present but induces a unique doomsday scenario in the future: model (i) leads to a BR; model (ii) leads to a LR; model (iii) leads to a LSBR. 
At late time, the parameter of EoS of DE for each of these models is very close to but slightly smaller than $-1$, thus corresponding to a phantom-like behaviour of DE. 
Despite the small variations of the parameter of the EoS for the three models, the asymptotic behaviour of the Universe is quite different from the one in $\Lambda$CDM, with the unavoidable rip of all the structure in the Universe no matter the interaction that bound them.

The cosmological parameters of the models are fixed as follows:  the value of $H_0$ and $\Omega_{\mat,0}$ in all the models, as well as of $w_\de$ in model (i) were fixed using the best fit in the Planck data \cite{wikiesa}. For the models (ii) and (iii) we fix the respective parameters $\LRparamA$ and $\LSBRparamA$ so that the amplitude of the DE energy density perturbations is the same for all models in the distant past. This choice of parameters was made so as to better understand the relative effects of each model on the evolution of the perturbations. In addition, we fix the effective squared speed of sound of the DE fluid, defined as $\delta p_\de/\delta \rho_\de$ in the rest frame of the fluid, to unity. This choice was made to remove potential instabilities in the dark sector that would lead the DE perturbations to quickly violate observational constraints.
 An improved analysis would incorporate in the calculations an observationally constrained value for  all the parameters , obtained by constraining the homogeneous and isotropic evolution of the model using standard candles like SNeIa.

For each of the models studied, we analyse the evolution of the linear cosmological perturbations  in absence of an anisotropic stress tensor and considering non-adiabatic contributions for the DE perturbations. In particular, we compute numerically the evolution of the matter density contrast for the DM and DE components, together with the evolution of the gravitational potential. 
The integrations are performed from well inside the radiation era till the far future. The outcome, which we present in Fig.~\ref{perturbationsgraphs}, shows that in all the models there is a very similar behaviour in the evolution of the perturbations. The largest difference until the present time seems to lie in the magnitude of DE perturbations:  even though the initial value of $\delta_\de$ is the same in all the models, we find that the more the background model resembles $\Lambda$CDM, the smaller $\delta_\de$ is at the present time. This effect is observed for all the scales. On the other hand, our numerical results indicate that in the future, when the perturbations of DE are the only sources of the gravitational potential, a change of the sign of $\Psi$ occurs on a scale-dependent order: it starts at large scales and progressively affects the smaller ones. We also find that this effect is model dependent, in the sense that it happens first in the model (i), which leads the Universe to a BR singularity, then in model (ii) which leads to a LR abrupt event, while in model (iii) which leads to a less virulent LSBR abrupt event, this effect happens later. We interpret this result as a change in the behaviour of gravity, which in the far future becomes repulsive and starts to rip structures apart, an effect that was already discussed in previous works \cite{Frampton:2011sp,Bouhmadi-Lopez:2014cca} on the same kind of phantom DE-fuelled abrupt events.

Using the results of the numerical integrations, we obtain for each model theoretical prediction for the matter power spectrum, as observed today, and the late-time evolution of the observable combination $f\sigma_8$. 
In all three models the deviations to the results of $\Lambda$CDM are within the $\lesssim1\%$ margin  for the matter power spectrum (cf. Fig.~\ref{diflambda}),  and within the $\lesssim0.3\%$ margin for $f\sigma8$ (cf. Fig.~\ref{growth}).
Comparing the results of the three models we find that the deviations to $\Lambda$CDM, are stronger for the model (i) that leads the Universe to a BR and weaker in the model (iii) that leads the Universe to a LSBR, while the model (ii) that induces a LR in the future has an intermediate behaviour. This suggests that these effects become more noticeable the more the background model deviates from $\Lambda$CDM at the present time. 

We compare the evolution of $f\sigma_8$ for low reshift ($z\lesssim1.4$) with the latest observations and find that, for all the models, the curve of $f\sigma_8$ is within the error bars for most data points.
We quantify the deviation from the observations by calculating the corresponding  reduced $\chi^{2}$, which allows us to make a preliminary comparison of the results for the three models. We find that all models are only slightly worse that $\Lambda$CDM at fitting the observational data. Here, we note that we do not perform a true statistical comparison of our models with observations, in particular we do not take into account the difference in number of parameters of the models. Nevertheless, since the three models analysed in this work have the same number of free parameters, the reduced $\chi^{2}$  allows us to state that model (i) provides the worse fit, while model (iii) seems to be of the three models analysed the one that best fits the observational data. We thus conclude that although the $\Lambda$CDM model gives the best fit to the observations, we cannot exclude other models like the ones analysed in this work. 

In general, the results of this paper suggest the possibility of finding imprints of a phantom DE which can be in agreement with the current observations. Nevertheless, the effects are small and a further examination is necessary to find an indication that could more clearly differentiate each model. We also stress that since we did not make an observational fit of these models (from a homogeneous and isotropic point of view), all our results should be taken as a guideline for a more accurate study that we hope to carry in the future.

While the classical cosmological pertubations of these models at first order are well defined, as we have shown, there are still some fundamental and intrinsic problems related to phantom dark energy models. In fact, as discussed in \cite{Carroll:2003st,Hsu:2004vr,Dubovsky:2005xd}, when a particle-physics description of the phantom dark energy is attempted, some instabilities may rise in theory due to higher order effects. In all three models presented in this work, this kind of effects can potentially become more problematic as the Universe evolves into one of the cosmological events considered, as the energy density becomes increasingly high. Though a thorough examination of these effects and the compatibility of the phantom dark energy models with a particle physics description is outside the scope of this paper, we will take this question into account in a future work.

\section*{Acknowledgments}

The authors are grateful to J.~Beltr\'{a}n Jim\'{e}nez, V.~Salzano and the Referees for enlightening comments on a previous version of the paper. 
The work of IA is supported by a Santander-Totta fellowship ``Bolsas de Investiga\c{c}{\~a}o Faculdade de Ci{\^e}ncias (UBI) -
Santander Totta''. The work of MBL is supported by the  Basque Foundation of Science Ikerbasque. MBL and JMorais wish to acknowledge the support from the Basque government Grant No. IT956-16 (Spain), FONDOS FEDER under grant FIS2014-57956-P (Spanish government). JMorais is also thankful to UPV/EHU for a PhD fellowship. This research work is supported by the grant UID/MAT/00212/2013.  The authors acknowledge the COST Action CA15117 (CANTATA).

\appendix

\section{Mapping to a phantom scalar field}
\label{Mapping to a phantom scalar field}

The phantom scalar field, which was first introduced in Ref.~\cite{Caldwell:1999ew} as a way of obtaining an equation of state with $w<-1$, can be seen as a minimally coupled scalar field with the sign of the kinetic term in the Lagrangian switched, i.e.:
\footnote{Here, we remind the reader that on this work we employ the metric signature $(-\,+\,+\,+)$. This is the reason that the sign of the kinetic term in the Lagrangian does not coincide with the one in \cite{Caldwell:1999ew}, which uses the signature $(+\,-\,-\,-)$.}
\begin{align}
	L_\varphi = \frac{1}{2} g^{\mu\nu}\partial_u\varphi\partial_\nu\varphi - V(\varphi)
	\,.
\end{align}
In this sense, the phantom scalar field can be viewed as a particular case of the $K$-essence models with $K=-X$ \cite{ArmendarizPicon:2000dh}.
The change of sign in the Lagrangian with regards to the usual scalar field is propagated to the energy density and pressure, which for the phantom field case in a FLRW background are given by the expressions:
\begin{align}
	\label{App_phantom_density}
	\begin{split}
	\rho_\varphi &= -\frac{1}{2}\dot\varphi^2 + V(\varphi)
	\,,
	\\
	p_\varphi &= -\frac{1}{2}\dot\varphi^2 - V(\varphi)
	\,.
	\end{split}
\end{align}
Likewise, the term of the equation of motion of the field that depends on the derivative of the potential suffers a change of sign:
\begin{align}
	\label{App_phantom_EoM}
	\ddot\varphi + 3H\dot\varphi - \frac{d V}{d\varphi} = 0
	\,.
\end{align}
From Eqs.~\eqref{App_phantom_density} we find that for positive valued potentials the parameter of EoS of the phantom scalar field, $w_\varphi=p_\varphi/\rho_\varphi$, can evolve dynamically in the interval $]-\infty,\,-1]$, while from the equation of motion \eqref{App_phantom_EoM} we find that the field tends to climb the potential towards a state of higher energy \cite{Caldwell:1999ew}.

In Section~\ref{review} we have reviewed three phantom DE models that lead to abrupt events in the future: model (i) leads to a BR; model (ii) leads to a LR; model (iii) leads to a LSBR. At the background level, these models can be mapped to a phantom scalar field through the identifications
\begin{align}
	\label{App_phantom_dotphi}
	\begin{split}
	\dot\varphi^2 &= -\left(1 + w_\de\right)\rho_\de
	\,,
	\\
	V(\varphi) &= \frac{1}{2}\left(1-w_\de\right)\rho_\de
	\,.
	\end{split}
\end{align}
Once the background evolution is imposed, the shape of the potential can be found by integrating the first equation of~\eqref{App_phantom_dotphi} 
\begin{align}
	\label{App_phantom_phi_of_t}
	\varphi(t)-\varphi(t_*) = \int_{t}^{t_*}d\tilde{t}\sqrt{-\left(1+w_\de\right)\rho_\de}
	\,,
\end{align}
inverting the expression obtained to find $\rho(\varphi)$ and finally plugging the result in the second equation of~\eqref{App_phantom_dotphi} we get $V(\varphi)$. For the three models at hand, we can solve \eqref{App_phantom_phi_of_t} analytically well inside the DE dominated regime, i.e.,  for $t\geq t_*$ we have $H^2\approx(\kappa^2/3)\rho_\de$, obtaining for each case the potential \cite{Hao:2003ww,Hao:2003th,Dabrowski:2006dd,Albarran:2015cda,Albarran:2016ewi}
\begin{align}
	\label{App_phantom_pot_solutions}
	\begin{split}
	 \textrm{(i)}\qquad&
	 V(\varphi) = \rho_{\de,*}
	 	 \dfrac{1-w_\de}{2}
	 	 \exp\left[\sqrt{-3(1+w_\de)}\kappa\left(\varphi-\varphi_*\right)\right]
	 \,,
	 \\
	 \textrm{(ii)}\qquad&
	 V(\varphi) = \rho_{\de,*}
	 	\left\{
	 		\left[ 1 + \dfrac{3}{4}\left(\dfrac{\Omega_\textrm{lr}}{\Omega_{\de,*}}\right)^{1/4}\kappa\left(\varphi-\varphi_*\right)\right]^2 
	 		+ \dfrac{1}{2}\left(\dfrac{\Omega_\textrm{lr}}{\Omega_{\de,*}}\right)^{1/2}
	 	\right\}
	 	\left[ 1 + \dfrac{3}{4}\left(\dfrac{\Omega_\textrm{lr}}{\Omega_{\de,*}}\right)^{1/4}\kappa\left(\varphi-\varphi_*\right)\right]^2
	 \,,
	 \\
	 \textrm{(iii)}\qquad&
	 V(\varphi) = \rho_{\de,*}
	 	\left\{
	 		\left[ 1 + \dfrac{1}{2}\left(\dfrac{\Omega_\textrm{lsbr}}{\Omega_{\de,*}}\right)^{1/2}\kappa\left(\varphi-\varphi_*\right)\right]^2 
	 		+ \dfrac{1}{6}\left(\dfrac{\Omega_\textrm{lsbr}}{\Omega_{\de,*}}\right)^{1/2}
	 	\right\}
	 \,.
	\end{split}
\end{align}

We now look at the evolution of the linear scalar perturbations of such models. In the phantom scalar field description, the evolution of the perturbation $\delta\varphi$ in the Newtonian gauge is given by \cite{Caldwell:1999ew,Hwang:2005hb}
\begin{align}
	\delta\varphi'' + 2\mathcal{H}\delta\varphi' + \left(k^2 - a^2\frac{d^2V}{d\varphi^2}\right)\delta\varphi 
	=
	2a^2\frac{d V}{d\varphi}\Phi
	+\varphi'
	\left(\Phi'-3\Psi'\right) 
	\,.
\end{align}
As long as the squared mass term $k^2 - a^2d^2V/d\varphi^2$ is positive, then the perturbation $\delta\varphi$ is free of instabilities \cite{Caldwell:1999ew}. However, for the three models considered in this work, the second derivative of the potential is always positive; in fact, from Eq.~\eqref{App_phantom_pot_solutions} we find that for the models (i) and (ii) $d^2V/d\varphi^2$ goes to infinity as the field climbs the potential, while for the model (iii) it is a positive constant. Therefore, in all cases as the Universe expands the effective mass of the perturbation $\delta\varphi$ becomes imaginary and the unstable regime sets in.

As discussed above, for the three models considered in this work there is an instability at the perturbative level that leads $\delta\varphi$ to extremely large values in the far future, as the Universe evolves towards a cosmological singularity or abrupt event. However, as we will see next, if we describe the linear perturbations of the scalar field from an hydrodynamical point of view, this instability is not displayed for $\delta_\varphi=\delta\rho_\varphi/\rho_\varphi$ or for the gravitational potential $\Psi$. Let us begin by reviewing the perturbations of the energy density, pressure and the peculiar velocity of the phantom scalar, which in the Newtonian gauge read
\begin{align}
	\label{App_phantom_mapping_pert}
\begin{split}
	\delta\rho_\varphi &= -\frac{1}{a^2} \varphi' \left({\delta\varphi}' - \varphi'\Phi\right) + \frac{dV}{d\varphi}\delta\varphi
	\,,
	\\
	\delta p_\varphi &= - \frac{1}{a^2}\varphi' \left({\delta\varphi}' - \varphi' \Phi\right) - \frac{dV}{d\varphi}\delta\varphi
	\,,
	\\
	v_\varphi &= -\frac{\delta\varphi}{\varphi'}
	\,.
\end{split}
\end{align}
From these equations we find that
\begin{align}
	\label{App_phantom_scalar_decomposition}
	\delta p_\varphi = \delta\rho_\varphi - 2\frac{dV}{d\varphi}\delta\varphi = \delta\rho_\varphi + 2\frac{dV}{d\varphi}\varphi' v_\varphi
	\,,
\end{align}
Since the adiabatic squared speed of sound is $c_{a\varphi}^2=\dot{p_\varphi}/\dot{\rho_\varphi} = 1 - 2a^2/(3\mathcal{H} \varphi')(dV/d\varphi)$ a comparison of Eq.~\eqref{App_phantom_scalar_decomposition} with \eqref{pnew} leads to the conclusion that at the perturbative level the phantom scalar field corresponds to a non-adiabatic fluid with effective speed of sound $c_{s\varphi}^2=1$. The same result can be reached by using the definitions \eqref{newvar0} and \eqref{App_phantom_mapping_pert} and noting that in the rest frame of the field $\delta\varphi=0$.

To show that the divergence of $\delta\varphi$ does not necessarily translate into a divergence of $\delta_\varphi$, we now look at the evolution of the gravitational potential $\Psi$. Since at late-time the contribution of matter and radiation to the potential is suppressed by the factors $\Omega_\mat$ and $\Omega_\rad$, respectively, then we can assume that in the future the potential is generated only by the DE perturbations. By combining Eqs.~\eqref{GTequations2}, \eqref{App_phantom_EoM}, \eqref{App_phantom_mapping_pert} and \eqref{App_phantom_scalar_decomposition} and changing from $\eta$-derivatives to $x$-derivatives 
\footnote{We remind the reader that by definition $x=\log(a/a_0)$.
},
we arrive at the following evolution equation for each Fourier mode
\begin{align}
	\label{App_eqmil}
	\Psi_{xx}
	+ \left(
		2
		+ \frac{\mathcal{H}'}{\mathcal{H}^2}
		- 2\frac{\varphi''}{\mathcal{H}\varphi'}
	\right)\Psi_x
	+ 2\left(
		 \frac{\mathcal{H}'}{\mathcal{H}^2}
		- \frac{\varphi''}{\mathcal{H}\varphi'}
		+ \frac{k^{2}}{\mathcal{H}^2}
	\right)\Psi
	&=0
	\,.
\end{align}
At this point, we note that $\mathcal{H}'/\mathcal{H}^2=-(1+3w)/2$ and $\varphi''/(\mathcal{H}\varphi')=-(1+3c_{a\varphi}^2)/2$ and that the term in $k^2$ can be dropped, since in a phantom dominated (or in general in an inflating) Universe all the modes end up outside of the Hubble horizon. We can then approximate \eqref{App_eqmil} as:
\begin{align}
	\Psi_{xx}
	+ \frac{1}{2}\left[
		2
		+3\left(1+c_{a\varphi}^2\right)
		+3\left(c_{a\varphi}^2 - w_\varphi\right)
	\right]\Psi_x
	+ 3\left(c_{a\varphi}^2 - w_\varphi\right)\Psi
	&\approx0
	\,.
\end{align}
Therefore, there is no instability at the level of the gravitational potential as long as the mass term in the previous equation is positive. To check whether or not this is the case in the models at hand, we use the relation $c_{a}^2=w+\rho(w'/\rho')$ and recall that for model (i) $w_\varphi'=0$ while for models (ii) and (iii) $w_\varphi$ is approaching $-1$ from below and $\rho_\varphi$ grows continuously as the Universe expands. Thus we find that in all three cases $\Psi$ remains bounded and slowly varying with $x$. Finally, from the perturbed Einstein equations \eqref{GTequations3n}, we conclude that $\delta_\varphi=-2\Psi$ is also bounded at late-time. This is the behaviour in the far future that we find in Figs.~\ref{perturbationsgraphs} and \ref{phiphiphi}.

\section{Decomposition of a non-adiabatic pressure}
\label{AppA}

Here, we review the derivation of Eq.~\eqref{pnew} which relates the pressure perturbation of a fluid $A$ with its relative energy density perturbation and peculiar velocity. We begin by separating the pressure perturbation into its adiabatic, $\delta p_\mathrm{aA}\equiv c_{aA}^2\delta\rho_A$, and non-adiabatic, $\delta p_\mathrm{naA}$, components \cite{Valiviita:2008iv}
\begin{align}
	\label{ad_nonad_decomposition}
	\delta p_A = c_{aA}^2\delta\rho_A + \delta p_\mathrm{naA}
	\,.
\end{align}
Under a general gauge transformation $(\eta,\,x^i)\rightarrow(\eta,\,x^i) + (\delta \eta,\,\partial^i \delta x)$ the energy density and pressure perturbations and the peculiar velocity in the final ${(2)}$ and initial ${(1)}$ gauges are related through \cite{Valiviita:2008iv}
\begin{align}
	\label{gauge_transformations}
	\delta \rho_A^{(2)} = \delta \rho_A^{(1)} - \rho_A'\delta\eta
	\,,
	\qquad
	\delta p_A^{(2)} = \delta p_A^{(1)} - c_{aA}^2\rho_A'\delta\eta
	\,,
	\qquad
	v_a^{(2)}+B^{(2)} = v_a^{(1)}+B^{(1)} + \delta\eta
	\,,
\end{align}
where $B$ is related to the metric perturbation $\delta g_{0i}$ and is set to zero in the Newtonian gauge \cite{Valiviita:2008iv}. 
From Eq.~\eqref{gauge_transformations} we arrive to the conclusion that the non-adiabatic term in Eq.~\eqref{ad_nonad_decomposition} is gauge invariant. Since by definition, we have $\delta p_A = c_{sA}^2 \delta \rho_A$ in the rest frame of the fluid $A$, we can write $\delta p_\mathrm{naA}$ as \cite{Bean:2003fb,Valiviita:2008iv}
\begin{align}
	\delta p_\mathrm{naA} = \left.\delta p_A\right|_\textrm{r.f.} - c_{aA}^2 \left.\delta\rho_A\right|_\textrm{r.f.} = \left(c_{sA}^2 - c_{aA}^2\right)\left.\delta \rho_A\right|_\textrm{r.f.}
	\,.
\end{align}
Next, we note that in order to change from the rest frame gauge (${v_a}|_\textrm{r.f.}=0$, $B|_\textrm{r.f.}=0$) to the Newtonian one ($B=0$) we have to set $v_a=\delta\eta$, cf. Eq.~\eqref{gauge_transformations}. Therefore, we find that in the Newtonian gauge the previous equation can be written as \cite{Bean:2003fb,Valiviita:2008iv}
\begin{align}
	\delta p_A &~= c_{aA}^2 \delta\rho_A + \left(c_{sA}^2 - c_{aA}^2\right)\left(\delta \rho_A + \rho_A' v_A\right)
	\nonumber\\
	&~= c_{sA}^2 \delta\rho_A - 3\mathcal{H}\left(c_{sA}^2 - c_{aA}^2\right)\left(1 + w_A\right)\rho_A v_A
	\,.
\end{align}

\section{Classical perturbations for a Universe containing an adiabatic fluid with a negative speed of sound}
\label{AppB}

On the following appendix, we remind why a universe filled with a dark energy fluid with a negative speed of sound, like it is the case of a phantom fluid with a constant equation of state, the linear perturbations of dark energy blow up.
We begin by recalling the evolution equations \eqref{seteq1ncor} for the pair $\delta_A$ and $v_A$ in Fourier space:
\begin{align}
	\label{A1}
	&\delta_{A}'=
	3\mathcal{H}\left(w_A-c_{sA}^2\right)\delta_A
	+\left(1+w_A\right)\left[9\mathcal{H}^2\left(c_{sA}^2-c_{aA}^2\right) + k^2\right]v_A
	+ 3\left(1+w_A\right)\Psi'
	\,,\\
	\label{A2}
&v_{A}'=\left(3c^2_{sA}-1\right)\mathcal{H}v_A-\frac{c^{2}_{sA}}{1+w_A}\delta_A - \Psi
	\,.
\end{align}
We can combine these equations into a single second order inhomogeneous differential equation for $\delta_A$ by differentiating \eqref{A1} and then using \eqref{A1} and \eqref{A2} to eliminate $v_A$ and $v_A'$
\begin{align}
	&\delta_{A}'' 
	+\left[
		\left(
			1
			-6w_A
			+3c_{aA}^2
		\right)
		- \frac{\mathcal{H} \left(c_{sA}^2 - c_{aA}^2\right)' + 2\mathcal{H}' \left(c_{sA}^2 - c_{aA}^2\right)}{\mathcal{H}^2 \left(c_{sA}^2 - c_{aA}^2\right) + \frac{k^2}{9}}
	\right]
	\mathcal{H}\delta_A'
	\nonumber\\&~
	-3\biggr[
		\mathcal{H}' \left(w_A - c_{sA}^2\right)
		+ \mathcal{H} \left(w_A - c_{sA}^2\right)'
		+ \mathcal{H}^2 \left( 1 - 3w_A\right)\left( w_A - c_{sA}^2 \right)
		+ 3\mathcal{H}^2 w_A \left( c_{aA}^2 - c_{sA}^2 \right)
	\biggr.
	\nonumber\\
	&\biggr.\phantom{-3\left[\right.}
		+ \mathcal{H}^2\left( c_{sA}^2 - w_A\right)\frac{\mathcal{H} \left(c_{sA}^2 - c_{aA}^2\right)' + 2\mathcal{H}' \left(c_{sA}^2 - c_{aA}^2\right)}{\mathcal{H}^2 \left(c_{sA}^2 - c_{aA}^2\right) + \frac{k^2}{9}}
		-\frac{k^2}{3}c_{sA}^2
	\biggr]\delta_A
	\nonumber\\
	&~=
	3\left(1+w_A\right)
	\left\{
	\Psi''
	+ \left[ 
		\left(1 - 3c^2_{sA}\right)
		- \frac{\mathcal{H} \left(c_{sA}^2 - c_{aA}^2\right)' + 2\mathcal{H}' \left(c_{sA}^2 - c_{aA}^2\right)}{\mathcal{H}^2 \left(c_{sA}^2 - c_{aA}^2\right) + \frac{k^2}{9}}
	\right] \mathcal{H}\Psi'
	- \left[3\mathcal{H}^2\left(c_{sA}^2-c_{aA}^2\right) + \frac{k^2}{3}\right] \Psi
	\right\}
	\,.
\end{align}
This is a wave equation with a variable damping term, $\lambda_A$, and mass, $m_A$, and with an external source, $\mathcal{S}_A$:
\begin{align}
	\delta_{A}'' + \lambda_A \delta_A' + m^2_A \delta_A = \mathcal{S}_A
	\,,
\end{align}
where
\begin{align}
\begin{split}
	\lambda_A &~= \left[
		\left(
			1
			-6w_A
			+3c_{aA}^2
		\right)
		- \frac{\mathcal{H} \left(c_{sA}^2 - c_{aA}^2\right)' + 2\mathcal{H}' \left(c_{sA}^2 - c_{aA}^2\right)}{\mathcal{H}^2 \left(c_{sA}^2 - c_{aA}^2\right) + \frac{k^2}{9}}
	\right]
	\mathcal{H}
	\nonumber\\
	m_A^2 &~= k^2 c^{2}_{sA}
	-3\Bigg[
		\mathcal{H}' \left(w_A - c_{sA}^2\right)
		+ \mathcal{H} \left(w_A - c_{sA}^2\right)'
		+ \mathcal{H}^2 \left( 1 - 3w_A\right)\left( w_A - c_{sA}^2 \right)
		+ 3\mathcal{H}^2 w_A \left( c_{aA}^2 - c_{sA}^2 \right)
	\Bigg.
	\nonumber\\
	&~\Bigg.\phantom{= k^2 c^{2}_{sA}-3\Bigg[\Bigg.}
		+ \mathcal{H}^2\left( c_{sA}^2 - w_A\right)\frac{\mathcal{H} \left(c_{sA}^2 - c_{aA}^2\right)' + 2\mathcal{H}' \left(c_{sA}^2 - c_{aA}^2\right)}{\mathcal{H}^2 \left(c_{sA}^2 - c_{aA}^2\right) + \frac{k^2}{9}}
	\Bigg]
	\nonumber\\
	\mathcal{S}_A &~= 3\left(1+w_A\right)
	\left\{
	\Psi''
	+ \left[ 
		\left(1 - 3c^2_{sA}\right)
		- \frac{\mathcal{H} \left(c_{sA}^2 - c_{aA}^2\right)' + 2\mathcal{H}' \left(c_{sA}^2 - c_{aA}^2\right)}{\mathcal{H}^2 \left(c_{sA}^2 - c_{aA}^2\right) + \frac{k^2}{9}}
	\right] \mathcal{H}\Psi'
	- \left[3\mathcal{H}^2\left(c_{sA}^2-c_{aA}^2\right) + \frac{k^2}{3}\right] \Psi
	\right\}
\end{split}
\end{align}
To leading order in $k^2$, we have $m_A^2\simeq k^2 c^{2}_{sA}$. Therefore, for sufficiently large $k$, the mass becomes imaginary if $c^{2}_{sA}<0$ and the solutions of the homogeneous equation comprise a decaying and a growing mode, with the latter leading to the instabilities. As such, barotropic fluids with negative adiabatic squared speed of sound, for which $c_{sA}^2=c_{aA}^2$, are stricken by instabilities at the linear level in perturbations. The growth of $\delta_A$ can be particularly intense during the matter era, when most relevant modes are inside the Hubble horizon, i.e., $k^2\gg\mathcal{H}^2$, and $\Psi\sim const$.

%

\end{document}